\documentclass[nofootinbib, twocolumn,secnumarabic,amssymb, nobibnotes, aps, pra,superscriptaddress,longbibliography]{revtex4-2}

\usepackage{amsmath}
\usepackage{longtable}
\usepackage{makecell}
\usepackage{graphicx}
\usepackage{booktabs}
\usepackage{url}
\usepackage[colorlinks,linkcolor=blue,anchorcolor=blue,citecolor=blue,urlcolor=blue,pdfpagelabels=false]{hyperref}
\usepackage{mathtools}
\usepackage{color}
\usepackage{multirow}

\graphicspath{{Figures/}}

\begin{document}

\title{Geometric Parametric Instability in Nonlinear Multipass Cells}

\author{Chao Mei}
\email{meichao@nbu.edu.cn}
\affiliation{Department of Physics, School of Physical Science and Technology, Ningbo University, Ningbo 315211, China}

\author{Junhong Yang}
\affiliation{KEY $\&$ Core Technology Innovation Institute of the Greater Bay Area, Guangzhou 510725, China}

\author{Tao Sun}
\affiliation{KEY $\&$ Core Technology Innovation Institute of the Greater Bay Area, Guangzhou 510725, China}

\author{Qian Gao}
\affiliation{KEY $\&$ Core Technology Innovation Institute of the Greater Bay Area, Guangzhou 510725, China}

\author{Jinhui Yuan}
\affiliation{State Key Laboratory of Information Photonics and Optical Communications, Beijing University of Posts and Telecommunications, Beijing 100876, China}

\author{Jintao Fan}
\affiliation{Ultrafast Laser Laboratory, Key Laboratory of Optoelectronic Information Science and Technology of Ministry of Education, School 
of Precision Instruments and Opto-electronics Engineering, Tianjin University, Tianjin 300072, China
}

\author{Peilong Yang}
\affiliation{Laboratory of Infrared Materials and Devices,
    The Research Institute of Advanced Technologies, Ningbo University, Ningbo 315211, China}

\author{Günter Steinmeyer}
\affiliation{Institut für Physik, Humboldt-Universität zu Berlin, Newtonstraße 15, 12489 Berlin, Germany}
\affiliation{Max Born Institute for Nonlinear Optics and Short Pulse Spectroscopy, Max-Born-Straße 2a, 12489 Berlin, Germany}

\date{\today}

\begin{abstract}
\noindent
Geometric parametric instability (GPI) is the resonant growth of
discrete spectral sidebands enabled by longitudinally periodic
multimode evolution and has been studied primarily in graded-index
fibers. Here we show theoretically and numerically that GPI can occur in gas-filled nonlinear multipass cells (MPCs). By mapping a mode-matched MPC onto an equivalent waveguide, we derive a
Floquet quasi-phase-matching condition governed by the single-pass
Gouy-phase imbalance of the signal--idler pair relative to the pump
pair. The theory predicts the small-signal gain and bandwidth. A
pump-depleted coupled-mode model (CMM) further relates the maximum converted fraction to the residual phase mismatch. The CMM predicts multiple geometrically tunable sideband pairs associated with different radial indices and Floquet orders. For argon at $5$~bar, varying the cavity geometry shifts the sideband detuning from approximately $96$ to $46$~THz when the $p_{\mathrm{s}}=1$, $h=0$ branch is considered. A truncated multimode generalized nonlinear Schr\"odinger equation (MMGNLSE) model is used for numerical simulations with a semiclassical stochastic seed corresponding to one photon per spectral mode. The MMGNLSE simulations reproduce the predicted sideband frequencies and reveal pump depletion and competition among the retained radial channels. GPI in MPCs may therefore limit spatial beam quality in nonlinear pulse compression while providing a tunable mechanism for broadband multicolor generation.
\end{abstract}

\maketitle

\section{Introduction}

\noindent
Nonlinear propagation in multimode systems supports a range of
self-organized spatiotemporal phenomena \cite{Wright2016,Wright2022-1,Wright2022-2,Sun2024}. A prominent example is geometric parametric instability (GPI), which was first observed in graded-index (GRIN) multimode fibers~\cite{Krupa2016}. In a parabolic-index GRIN fiber, the nearly uniform spacing of the modal propagation constants produces periodic spatial self-imaging through the Talbot effect~\cite{Patorski1989,Wen2013}. The resulting beam breathing acts as a longitudinal parametric grating that compensates
the material-dispersion mismatch and drives the growth of discrete
spatiotemporal sidebands. Related multimode nonlinear dynamics in GRIN fibers including beam self-cleaning and broadband supercontinuum
generation have also been reported \cite{Lopez-Aviles2018,Arabi2018,Wright2015,Wright2017-1,Krupa2019}.

In parallel, gas-filled and solid-state MPCs have become
an important platform to study nonlinear effects, especially nonlinear pulse compression. By distributing the required nonlinear phase over multiple free-space passes, MPCs reduce the per-pass interaction strength to mitigate the possible material damage and ionization constraints \cite{Schulte2016,Hanna2021,Viotti2022}. At high peak or average power, however, Kerr self-focusing, diffraction, dispersion, and thermal lensing can distort the fundamental Gaussian mode and populate higher-order transverse modes~\cite{Daher2020,Yu2025}. The natural eigenmodes of a rotationally symmetric MPC are Laguerre--Gauss (LG) modes, whose single-pass Gouy phases are proportional to their transverse mode orders~\cite{Hanna2025}. The resulting equally spaced effective propagation constants resemble the modal spectrum responsible for GPI in parabolic-index GRIN fibers.

Quasi-phase-matched FWM and, more recently, vector modulation
instability have been investigated in gas-filled
MPCs~\cite{Hanna2020QPM,Escoto2026}. However, to the best of our knowledge, transverse-mode GPI driven by the Gouy-phase imbalance has not previously been investigated. Determining whether it can occur is relevant both as a possible limitation and as a potential functionality of MPCs. Uncontrolled sideband growth could degrade spatial mode quality and pulse-compression fidelity, whereas selective excitation of higher-order modes could enable spatially structured multicolor generation \cite{Ierano2026,Wei2023,Meena2023}. The accessible spatial channels are further constrained by orbital-angular-momentum conservation, which permits a Gaussian pump to generate counter-rotating vortex sidebands~\cite{Offer2021}.

In this work, we develop an analytical and numerical framework for GPI in nonlinear MPCs by treating each mirror-to-mirror pass as one period of a Floquet map. The resulting quasi-phase-matching condition is governed by the modal Gouy-phase imbalance accumulated over the single-pass length $L_\mathrm{cav}$ of the MPC cavity, rather than by the continuous self-imaging modulation of a GRIN fiber. We derive the sideband frequencies together with the small-signal gain
and bandwidth. Pump depletion is described using a reduced coupled-mode model (CMM). The analytical predictions are then tested using truncated multimode generalized nonlinear Schr\"odinger equation (MMGNLSE) simulations. The framework further shows how the Floquet sidebands can be tuned through the cavity geometry and gas pressure.

\section{Theoretical Framework}
\label{sec:theory}
\noindent This section develops the theoretical framework in four steps: the effective modal structure of the MPC, the nonlinear coupling and spatial-mode selection rules, the Floquet QPM condition, and the resulting small-signal gain and pump-depleted dynamics. To this end, we consider a simplified collinear equivalent-waveguide model of a symmetric Herriott-type MPC~\cite{Herriott1964}. The nonlinear phase is retained during every mirror-to-mirror pass, and the resulting parametric coupling accumulates over successive passes. However, the model neglects direct nonlinear interaction between subsequent passes at their crossing point inside the cell. This approximation is justified when the pulse duration is much shorter than the temporal separation between successive passages through a crossing region. As we also neglect astigmatism, the validity of our model is restricted to rotationally symmetric MPC geometries with ring-like spot patterns~\cite{Kong2020,Wang2024}. 

\subsection{Effective Modal Structure of MPCs: The Equivalent Waveguide}
\label{sec:modal}

\noindent We consider a symmetric MPC formed by two identical spherical mirrors with radius of curvature $R_0$ and mechanical mirror separation $L_\mathrm{cav}$. The corresponding round-trip length is $L_\mathrm{rt}=2L_\mathrm{cav}$. The free spectral range is
$\mathrm{FSR}=c/L_\mathrm{rt}=c/(2L_\mathrm{cav})$. In the present symmetric and radially resolved model, we use one single pass of length $L_\mathrm{cav}$, rather than one round trip, as the elementary step of the Floquet map.

For a cylindrically symmetric paraxial optical system, a convenient orthonormal basis is provided by the Laguerre--Gaussian (LG) modes~\cite{Haus1984,SalehTeich2007}:
\begin{equation}
    \begin{split}
    \mathrm{LG}_{\ell p}
    &= \sqrt{\frac{2}{\pi}}\frac{\mathcal{C}_{\ell p}}{w(\zeta)}
    \left(\frac{\sqrt{2}\,r}{w(\zeta)}\right)^{|\ell|}
    L_p^{|\ell|}\!\left(\frac{2r^2}{w^2(\zeta)}\right)
    \\ 
    &\times \exp\!\left[-\frac{r^2}{w^2(\zeta)}
    -i\!\left(\frac{\beta_0 r^2}{2R(\zeta)}+\ell\theta-\psi_{\ell p}(\zeta)\right)\right].
    \label{eq:LG_mode}
\end{split}
\end{equation}
Here $\zeta$ is the local longitudinal coordinate with $-L_\mathrm{cav}/2\leq\zeta\leq L_\mathrm{cav}/2$, and the mirrors are located at $\zeta=\pm L_\mathrm{cav}/2$. The quantity $\beta_0=\beta(\omega_0)$ is the propagation constant at the pump angular frequency $\omega_0$. Both $\ell$ and $p$ are integers ($p\geq0$) that denote the azimuthal and radial indices, respectively; $L_p^{|\ell|}$ is a generalized Laguerre polynomial, and $\mathcal{C}_{\ell p}=\sqrt{p!/(p+|\ell|)!}$ is the normalization constant. 

Let $\lambda_0=2\pi c/\omega_0$ be the vacuum pump wavelength, where
$c$ is the speed of light in vacuum, and $\omega_0$ is the pump angular frequency. For a waist radius $w_0$, the Rayleigh range is $z_R=\pi w_0^2/\lambda_0$. The beam radius and
wavefront curvature are $w^2(\zeta)=w_0^2(1+\zeta^2/z_R^2)$ and
$R(\zeta)=\zeta(1+z_R^2/\zeta^2)$, respectively. The Gouy phase of mode $\mathrm{LG}_{\ell p}$ is $\psi_{\ell p}(\zeta) = N_{\ell p}\arctan(\zeta/z_R)$, with $N_{\ell p}=1+2p+|\ell|$. The quantity $N_{\ell p}$ is the transverse-mode-order multiplier of the Gouy phase. Modes with the same $2p+|\ell|$ belong to the same degenerate family. With the waist at $\zeta=0$, the mirrors lie at $\zeta=\pm L_\mathrm{cav}/2$. Mode matching requires
$|R(\pm L_\mathrm{cav}/2)|=R_0$.

The single-pass Gouy phase of the fundamental LG$_{00}$ mode is \cite{Hanna2025}
\begin{equation}
    \Phi_0(C)
    =2\arctan\!\left[
      \frac{C}{\sqrt{C(2-C)}}
    \right],
    \qquad
    C=\frac{L_\mathrm{cav}}{R_0}.
    \label{eq:Phi0}
\end{equation}
For the symmetric two-mirror geometry considered here, the standard
resonator stability condition $|1-L_\mathrm{cav}/R_0|<1$ gives
$0<C<2$~\cite{Siegman1986}. The confocal geometry corresponds to
$C=1$. During one mirror-to-mirror pass, mode $\mathrm{LG}_{\ell p}$
accumulates the Gouy phase $\Phi_{\ell p}=N_{\ell p}\Phi_0$. For fixed $\ell$, adjacent radial modes satisfy $N_{\ell,p+1}-N_{\ell p}=2$ and are therefore separated by the single-pass Gouy phase $2\Phi_0$. This uniform spacing is the MPC analog of the equally spaced modal propagation constants in an ideal parabolic-index GRIN fiber.

In the equivalent-waveguide representation, the mode-dependent Gouy
phase can be expressed as an effective propagation constant,
\begin{equation}
  \widetilde{\beta}_{\ell p}
  =
  \frac{N_{\ell p}\Phi_0}{L_\mathrm{cav}}.
  \label{eq:effective_Gouy_propagation_constant}
\end{equation}
The tilde distinguishes this effective geometric contribution from the material propagation constant $\beta(\omega)$. Adjacent radial modes are consequently separated by $2\Phi_0/L_\mathrm{cav}$ in the effective propagation constant. The fundamental mode accumulates the Gouy phase $\Phi_0$ during each mirror-to-mirror pass. Thus, $N_\mathrm{SI}=2\pi/\Phi_0$ is the equivalent number of passes associated with a total Gouy-phase accumulation of $2\pi$, and
$L_\mathrm{SI}=N_\mathrm{SI}L_\mathrm{cav}$ is the corresponding
characteristic self-imaging length.

We denote the actual number of completed mirror-to-mirror passes by
the nonnegative integer $J$ and the corresponding accumulated
propagation distance by
\begin{equation}
  Z_J=JL_\mathrm{cav}.
  \label{eq:propagation_distance}
\end{equation}
For the full LG mode set, exact field reimaging occurs when
$J\Phi_0=2\pi m$ for some integer $m$. Within the purely radial $\ell=0$ subspace, whose adjacent modes differ
by $2\Phi_0$, a radial recurrence can occur when
$2J\Phi_0=2\pi m$. Unlike in a GRIN fiber, $L_\mathrm{SI}$ is not the elementary Floquet period of the MPC. The same mirror-to-mirror propagation map repeats after every distance $L_\mathrm{cav}$. The Floquet reciprocal vectors are therefore $2\pi h/L_\mathrm{cav}$, where $h\in\mathbb{Z}$ is the Floquet order. The Kerr interaction acts continuously within each pass, but its parametric effect is governed by the pump--sideband Gouy-phase imbalance accumulated over successive passes. 

\begin{figure}[h]
  \centering
  \includegraphics[width=0.48\textwidth]{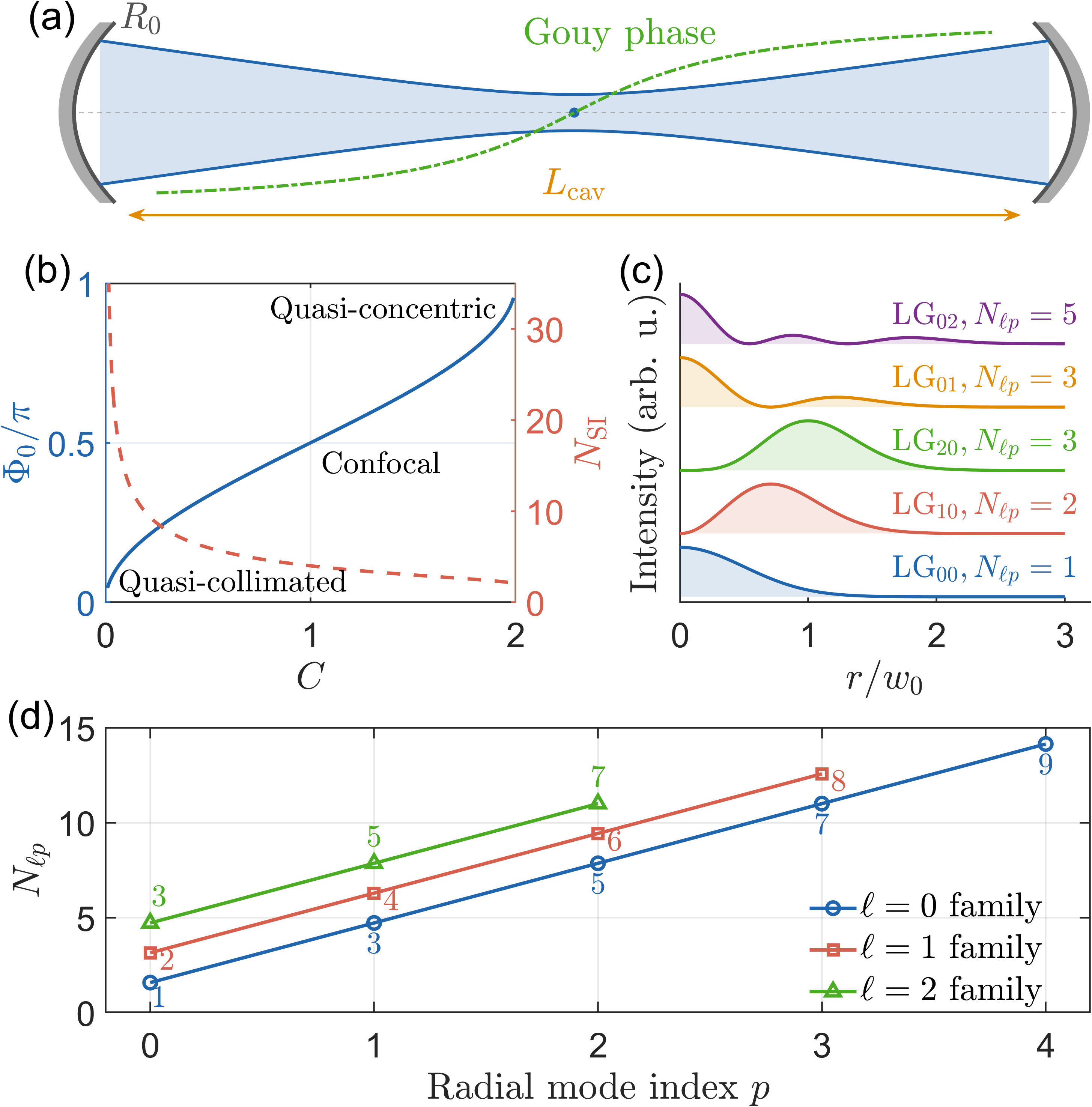}
  \caption{Effective modal structure of the symmetric MPC. (a) Mode-matched cavity and the Gouy phase accumulated during one mirror-to-mirror pass. (b) Fundamental-mode single-pass Gouy phase $\Phi_0/\pi$ and characteristic self-imaging pass number $N_\mathrm{SI}$ versus $C=L_\mathrm{cav}/R_0$, spanning the quasi-collimated, confocal, and quasi-concentric regimes. (c) Radial intensity profiles of representative LG modes. (d) Gouy-phase multiplier $N_{\ell p}=1+2p+|\ell|$ versus radial order $p$ for $\ell=0,1,2$. Adjacent radial orders at fixed $\ell$ differ by $\Delta N_{\ell p}=2$.}
  \label{fig:modes}
\end{figure}

Figure~\ref{fig:modes} connects the physical MPC geometry to its
effective modal structure. The symmetric mirror-to-mirror geometry and the Gouy phase accumulated during one pass are illustrated [Figure~\ref{fig:modes}(a)]. The increasing $C$ from the
quasi-collimated to the quasi-concentric limit increases the
single-pass Gouy phase $\Phi_0$ and correspondingly decreases the
characteristic self-imaging pass number $N_\mathrm{SI}$ [Figure~\ref{fig:modes}(b)]. At the confocal point $C=1$, $\Phi_0=\pi/2$ and $N_\mathrm{SI}=4$. Figure~\ref{fig:modes}(c) illustrates the increasingly structured radial intensity profiles of the LG modes. The $N_{\ell p}$ increases linearly with radial order $p$, with $N_{\ell,p+1}-N_{\ell p}=2$ at fixed $\ell$ [Figure~\ref{fig:modes}(d)]. This uniform mode-order spacing produces the equally spaced Gouy phases that underlie the Floquet phase-matching condition derived below.

\subsection{Nonlinear Coupling and Selection Rules}
\label{sec:coupling}

\noindent The strength of the nonlinear coupling between the LG modes depends on their transverse spatial overlap. We normalize the integration of each LG mode to unit scale, $\iint|\mathrm{LG}_{\ell p}|^2r\,\mathrm{d}r\,\mathrm{d}\theta=1$. For compactness, we introduce the composite index $\mu_j=(\ell_j,p_j)$ where $j=0,1,2,3$, together with the notation $\mathrm{LG}_{\mu_j}=\mathrm{LG}_{\ell_jp_j}$ and $N_{\mu_j}=1+2p_j+|\ell_j|$. The four-mode overlap tensor at the beam waist is defined as~\cite{Offer2021}
\begin{equation}
  S_{\mu_0\mu_1\mu_2\mu_3}
  =
  \pi w_0^2
  \iint
  \mathrm{LG}_{\mu_0}^{*}
  \mathrm{LG}_{\mu_1}
  \mathrm{LG}_{\mu_2}
  \mathrm{LG}_{\mu_3}^{*}
  r\,\mathrm{d}r\,\mathrm{d}\theta .
  \label{eq:S_tensor}
\end{equation}
The prefactor $\pi w_0^2$ makes the overlap dimensionless and gives
$S_{0,0,0,0}=1$ for the fundamental Gaussian mode. With this index ordering, the nonlinear term driving
$A_{\mu_0}$ is proportional to $S_{\mu_0\mu_1\mu_2\mu_3}
A_{\mu_1}A_{\mu_2}A_{\mu_3}^{*}$. Thus, $\mu_0$ labels the mode onto which the nonlinear polarization is projected, whereas $\mu_3$ labels the complex-conjugated partner. For the degenerate FWM driving the signal mode, $(\mu_0,\mu_1,\mu_2,\mu_3) = (\mu_\mathrm{s},\mu_\mathrm{p},\mu_\mathrm{p},\mu_\mathrm{i})$, so that the corresponding field product is $A_\mathrm{p}^{2}A_\mathrm{i}^{*}$. The analogous term driving the
idler is obtained by interchanging $\mu_\mathrm{s}$ and $\mu_\mathrm{i}$. The azimuthal integral in Eq.~\eqref{eq:S_tensor} is nonzero only when $\ell_1+\ell_2=\ell_0+\ell_3$. For degenerate FWM, this condition becomes $2\ell_\mathrm{p}=\ell_\mathrm{s}+\ell_\mathrm{i}$, which expresses conservation of orbital angular momentum. For a Gaussian pump ($\ell_\mathrm{p}=0$), the signal and idler can carry opposite topological charges, $\ell_\mathrm{i}=-\ell_\mathrm{s}$.

The Gouy-phase contribution to the FWM phase mismatch originates from
the difference between the combined mode orders of the signal--idler
pair and the two pump fields~\cite{Hanna2025}. We therefore define the dimensionless mode-order multiplier imbalance as
\begin{equation}
  \Delta N
  =
  N_{\mu_\mathrm{s}}
  +
  N_{\mu_\mathrm{i}}
  -
  2N_{\mu_\mathrm{p}}
  =
  4p_\mathrm{s},
  \label{eq:DeltaNG}
\end{equation}
where the final equality holds when the purely radial branch is driven by a fundamental Gaussian pump: $\mu_\mathrm{p}=(0,0)$ and $\mu_\mathrm{s}=\mu_\mathrm{i}=(0,p_\mathrm{s})$. The corresponding imbalance phase accumulated during one mirror-to-mirror pass is $\Delta N \Phi_0=4p_\mathrm{s}\Phi_0$. For $p_\mathrm{s}=0$, $\Delta N$ vanishes and the branch corresponds
to intramodal QPM-FWM in the fundamental spatial mode, as previously
investigated in MPCs~\cite{Hanna2020QPM,Escoto2026}. The spatial GPI
branches considered here instead require $p_\mathrm{s}\geq1$.

The SPM-induced nonlinear coefficient at the beam waist for fundamental mode is calculated as $\gamma_0=n_2\omega_0/(c\pi w_0^2)$ by using Eq.~\eqref{eq:S_tensor}. Here $n_2$ is the nonlinear refractive index of the gas in the MPC. Because the beam radius varies within a mirror-to-mirror pass, the local coefficient is $\gamma(\zeta) =\gamma_0w_0^2/w^2(\zeta) =\gamma_0/(1+\zeta^2/z_R^2)$. The longitudinal dependence of $\gamma(\zeta)$ is represented by its single-pass average,
\begin{equation}
  \gamma_\mathrm{eff}(C)
  =
  \frac{1}{L_\mathrm{cav}}
  \int_{-L_\mathrm{cav}/2}^{L_\mathrm{cav}/2}
  \gamma(\zeta)\,\mathrm{d}\zeta
  =
  \gamma_0F(C).
  \label{eq:gamma_eff}
\end{equation}
where 
\begin{equation}
    F(C) = \frac{\Phi_0(C)}{2}\sqrt{\frac{(2-C)}{C}}
\end{equation}
follows from the symmetric-cavity relation $2z_R/L_\mathrm{cav}=\sqrt{(2-C)/C}$.

For the radially symmetric modal branch, we use $S_{p_0,p_1,p_2,p_3} = S_{\mu_0\mu_1\mu_2\mu_3}$ with $\mu_j=(0,p_j)$. The three nonlinear processes retained in the GPI model are pump SPM, pump-induced cross-phase modulation (XPM) of the sidebands, and phase-sensitive FWM. Their effective nonlinear coefficients are defined as
$\gamma_\mathrm{SPM} = \gamma_\mathrm{eff}S_{0,0,0,0}$, $\gamma_\mathrm{XPM}(p_\mathrm{s}) = 2\gamma_\mathrm{eff} S_{p_\mathrm{s},0,0,p_\mathrm{s}}$ and $\gamma_\mathrm{FWM}(p_\mathrm{s}) = \gamma_\mathrm{eff} S_{p_\mathrm{s},0,0,p_\mathrm{s}}$. Consequently, $\gamma_\mathrm{XPM}(p_\mathrm{s})
= 2\gamma_\mathrm{FWM}(p_\mathrm{s})$. For $p_\mathrm{s}=1$, the overlap $S_{1,0,0,1}=1/2$ gives $\gamma_\mathrm{FWM}(1) = \gamma_\mathrm{eff}/2$ and $\gamma_\mathrm{XPM}(1) = \gamma_\mathrm{SPM}= \gamma_\mathrm{eff}$.
 
\subsection{Floquet Quasi-Phase-Matching Condition}
\label{sec:QPM}
 
\begin{figure}[h]
  \centering
  \includegraphics[width=0.48\textwidth]{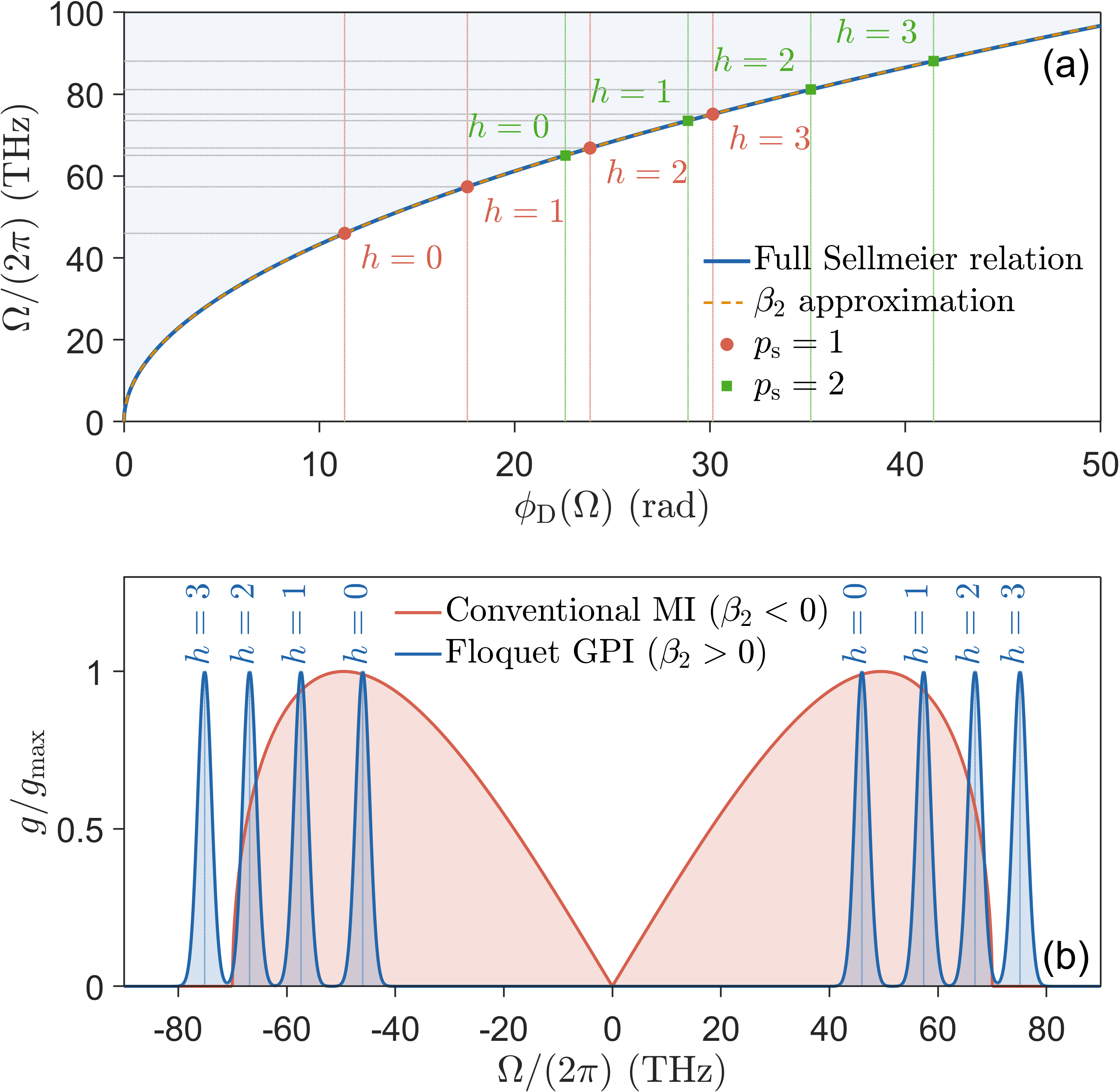}
  \caption{Floquet phase-matching construction for a quasi-concentric MPC with $C=1.95$, filled with argon at $5$~bar and pumped at $\lambda_0=1030$~nm. (a) Dispersion relation $\Omega(k_\mathrm{D})$ with $k_\mathrm{D}=\phi_\mathrm{D}/L_\mathrm{cav}$ and the accumulated phase $\phi_\mathrm{D}$ during one mirror-to-mirror pass. The blue curve uses the full argon Sellmeier relation, and the overlying orange dashed curve shows a $\beta_2$-dominated approximation. Vertical dotted lines denote the Floquet branches for $p_\mathrm{s}=1,2$ and
$h=0,1,2,3$. Red circles and green squares mark the corresponding
intersections, and horizontal dotted lines project their detunings.
(b) Schematic comparison of the discrete GPI bands in normal
dispersion (blue) with the scalar MI gain region in anomalous dispersion (red).}
  \label{fig:mechanism}
\end{figure}

\noindent In the coupled-mode description, material dispersion, Gouy phases, SPM, and XPM determine the residual phase mismatch, whereas the FWM term drives energy transfer. For a degenerate FWM process, the signal and idler are symmetrically detuned from the pump
and have angular frequencies $\omega_0+\Omega$ and
$\omega_0-\Omega$, respectively. Their symmetric material-dispersion phase relative to two pump photons accumulated during one pass is
\begin{equation}
  \phi_\mathrm{D}(\Omega)=\left[\beta(\omega_0+\Omega)+\beta(\omega_0-\Omega)-2\beta(\omega_0)\right]
  L_\mathrm{cav}.
  \label{eq:disp_phase}
\end{equation}
Defining $k_\mathrm{D}(\Omega)=\phi_\mathrm{D}(\Omega)/L_\mathrm{cav}$,
the inverse relation $\Omega(k_\mathrm{D})$ gives the dispersion curve
used for the phase-matching construction in Fig.~\ref{fig:mechanism}(a).
For the broadband sidebands, we evaluate the full propagation constant $\beta(\omega)=\omega n(\omega)/c$, with $n(\omega)$ obtained from the Sellmeier relation for argon gas~\cite{Dalgarno1960,Peck1964,Borzsonyi2008}. The nonlinear mismatch arises from the difference between the pump SPM phase and the pump-induced XPM phases of the sidebands. Because the signal and idler acquire equal XPM shifts, the single-pass nonlinear contribution is 
\begin{equation}
\Delta\phi_\mathrm{NL}(p_\mathrm{s})=2\left[\gamma_\mathrm{SPM}-\gamma_\mathrm{XPM}(p_\mathrm{s})\right] P_\mathrm{p}L_\mathrm{cav}
\label{eq:nonlinear_phase}
\end{equation}
For the $p_\mathrm{s}=1$ sideband mode, $\gamma_\mathrm{XPM}(1) =\gamma_\mathrm{SPM}$, leading to $\Delta\phi_\mathrm{NL}(1)=0$. This cancellation concerns only the net nonlinear phase mismatch. The individual pump-SPM and sideband-XPM phase shifts remain nonzero.

Because phase is defined modulo $2\pi$, the residual phase mismatch for Floquet order $h$ is
\begin{equation}
  \delta\phi(p_\mathrm{s},h,\Omega)
  =
  \phi_\mathrm{D}(\Omega)
  -
  \Delta N \Phi_0
  +
  \Delta\phi_\mathrm{NL}(p_\mathrm{s})
  -
  2\pi h,
  \label{eq:residual_mismatch}
\end{equation}
The Floquet phase-matching condition $\delta\phi(p_\mathrm{s},h,\Omega)=0$ leads
\begin{equation}
  \phi_\mathrm{D}(\Omega)
  =
  \Delta N \Phi_0
  -
  \Delta\phi_\mathrm{NL}(p_\mathrm{s})
  +
  2\pi h.
  \label{eq:Floquet_QPM}
\end{equation}
In a uniform Kerr medium, conventional scalar modulation instability (MI) typically requires anomalous dispersion \cite{Agrawal2019}. In contrast, Floquet phase matching can occur in the normal dispersion regime. To visualize this condition, Fig.~\ref{fig:mechanism}(a) plots the positive frequency detuning $\Omega/(2\pi)$ as a function of the dispersive phase $\phi_\mathrm{D}(\Omega)$. Expanding the propagation constant about the pump frequency gives
\begin{equation}
  \frac{\phi_\mathrm{D}(\Omega)}{L_\mathrm{cav}}
  =
  \beta_2\Omega^2
  +
  \frac{\beta_4}{12}\Omega^4
  +\cdots ,
  \label{eq:disp_phase_expansion}
\end{equation}
where $\beta_m = \left. \mathrm{d}^m\beta/\mathrm{d}\omega^m \right|_{\omega_0}$ is the $m$th-order dispersion coefficient. All odd-order terms cancel because the signal and idler are symmetrically detuned about the pump. 

The full argon Sellmeier result and the leading-order approximation $\phi_\mathrm{D}(\Omega)\simeq\beta_2\Omega^2L_\mathrm{cav}$ are compared in Fig.~\ref{fig:mechanism}(a). Because the nonlinear phase-mismatch contribution $\Delta\phi_\mathrm{NL}$ is set to zero,
the vertical line for each $(p_\mathrm{s},h)$ branch is located at
the phase value $\phi_\mathrm{D}=4p_\mathrm{s}\Phi_0+2\pi h$.
Its intersection with the curve of $\phi_\mathrm{D}(\Omega)$ determines the corresponding phase-matched sideband detuning
$\Delta f_\mathrm{SB}(p_\mathrm{s},h)$. The right-hand side of Eq.~\eqref{eq:Floquet_QPM} changes by $2\pi$ when the Floquet order increases from $h$ to $h+1$. The resulting intersection with $\phi_\mathrm{D}(\Omega)$ therefore occurs at a different detuning, producing another pair of Floquet sidebands. The sideband positions are determined by material dispersion, the modal Gouy-phase imbalance, and the Floquet order. Figure~\ref{fig:mechanism}(b) compares the qualitative gain structures of conventional scalar MI and Floquet GPI. In anomalous dispersion, scalar MI is supported over a continuous detuning interval determined by the balance between Kerr nonlinearity and material dispersion \cite{Agrawal2019}. In the normally dispersive MPC, by contrast, the integer Floquet order $h$ provides discrete phase-matching conditions, producing separated gain bands centered at the intersections identified in Fig.~\ref{fig:mechanism}(a). The curves in panel (b) are normalized independently and illustrate only the different spectral structures.

For a sideband detuning $\Delta f_{\mathrm{SB}}(p_\mathrm{s},h)$, the corresponding signal and idler frequencies are $f_0+\Delta f_{\mathrm{SB}}$ and $f_0-\Delta f_{\mathrm{SB}}$, respectively, where $f_0=\omega_0/2\pi$ is the pump frequency. Neglecting the nonlinear phase mismatch and higher-order dispersion,
and substituting $\phi_\mathrm{D}(\Omega)\simeq\beta_2\Omega^2L_\mathrm{cav}$
into Eq.~\eqref{eq:Floquet_QPM}, we obtain
\begin{equation}
  \Delta f_{\mathrm{SB}}(p_\mathrm{s},h)
  \approx
  \frac{1}{2\pi}
  \sqrt{
    \frac{
      4p_\mathrm{s}\Phi_0+2\pi h
    }{
      \beta_2L_\mathrm{cav}
    }
  }.
  \label{eq:sideband_floquet}
\end{equation}
For $\beta_2>0$, this approximation yields a real nonzero sideband detuning only when $4p_\mathrm{s}\Phi_0+2\pi h>0$. Floquet orders that do not satisfy this condition have no positive-detuning phase-matched solution. For the $h=0$ branch, Eq.~\eqref{eq:sideband_floquet} provides the approximate geometric scaling $\Delta f_{\mathrm{SB}}(p_\mathrm{s},0) \propto\sqrt{\Phi_0/L_\mathrm{cav}}$, whereas the $h\neq0$ branches additionally depend on the Floquet phase $2\pi h$.
  
\subsection{GPI Gain and Pump-Depleted Dynamics}
\label{sec:gain}

\noindent Having established the Floquet QPM condition, we next consider the small-signal gain and pump-depleted dynamics. Detailed derivations of the pump-depleted and small-signal
coupled-mode models are given in Appendices~\ref{app:pump_depleted_model} and \ref{app:small_signal_details}, respectively.

\subsubsection{Small-Signal Regime}
\label{subsec:small_signal}

\noindent In the small-signal regime under the undepleted-pump approximation, the pump power $P_\mathrm{p}$ is treated as a constant. The effective FWM nonlinear coefficient $\gamma_\mathrm{FWM}$ defined above yields the phase-sensitive coupling strength
\begin{equation}
  \kappa(p_\mathrm{s})
  =
  \gamma_\mathrm{FWM}(p_\mathrm{s})P_\mathrm{p},
  \label{eq:FWM_coupling}
\end{equation}
where the quantity $\kappa$ has units of inverse length and describes the signal--idler coupling per unit propagation distance. In order to calculate the small-signal amplitude gain $g$, we linearize the coupled signal--idler equations under the undepleted-pump approximation. The resulting matrix system and eigenvalue relation are given by Eqs.~\eqref{eq:small_signal_matrix} and \eqref{eq:gain_eigenvalue}, respectively, in Appendix~\ref{app:small_signal_details}. The positive branch of the solution is
\begin{equation}
  g(p_\mathrm{s},h,\Omega)
  =
  \sqrt{
    |\kappa(p_\mathrm{s})|^2
    -
    \left[
      \frac{\delta\phi(p_\mathrm{s},h,\Omega)}
           {2L_\mathrm{cav}}
    \right]^2
  }.
  \label{eq:gain}
\end{equation}
Exponential gain occurs when the radicand in Eq.~\eqref{eq:gain} is positive, namely,
\begin{equation}
  \left|
    \delta\phi(p_\mathrm{s},h,\Omega)
  \right|
  <
  2
  \left|
    \kappa(p_\mathrm{s})
  \right|
  L_\mathrm{cav}.
  \label{eq:gain_condition}
\end{equation}
Outside this interval, $g$ is imaginary, and the evolution is
oscillatory rather than exponentially growing. At exact Floquet phase
matching $\delta\phi(p_\mathrm{s},h,\Omega)=0$, the maximum value of $g(p_\mathrm{s},h,\Omega)$ is $g_\mathrm{max}(p_\mathrm{s},h,\Omega)=|\kappa(p_\mathrm{s})|$. Thus, for a fixed $p_\mathrm{s}$, all $h$ have the same peak gain $|\kappa(p_\mathrm{s})|$. However, according to Eq.~\eqref{eq:sideband_floquet}, the center frequencies of the sidebands $\Delta f_{\mathrm{SB}}(p_\mathrm{s},h)$ are different.

It has been reported that the standard undepleted-pump coupled-mode solution exhibits hyperbolic transfer coefficients
$\cosh(|\kappa|Z_J)$ and $\sinh(|\kappa|Z_J)$
\cite{Agrawal2019,Boyd2020}. At exact Floquet phase matching, ensemble averaging over statistically independent signal and idler seeds eliminates the phase-sensitive cross terms, yielding the relative sideband power gain
\begin{equation}
  \mathcal{G}(p_\mathrm{s},J)=
  \cosh\!\left(
    2J|\kappa(p_\mathrm{s})|L_\mathrm{cav}
  \right).
  \label{eq:noise_seeded_gain_exact}
\end{equation}
Here $g$ is the amplitude-growth rate per unit length, whereas
$\mathcal{G}$ is the relative sideband power gain after $J$ passes.
The finite-mismatch expression is given by Eq.~\eqref{eq:noise_seeded_power_gain} in Appendix~\ref{app:small_signal_details}.

Near the phase-matched frequency $\Omega_\mathrm{SB}=2\pi\Delta f_\mathrm{SB}(p_\mathrm{s},h)$, the residual mismatch can be linearized as
\begin{equation}
  \delta\phi(p_\mathrm{s},h,\Omega)
  \simeq
  \left.
    \frac{\mathrm{d}\phi_\mathrm{D}}
         {\mathrm{d}\Omega}
  \right|_{
    \Omega=
    2\pi \Delta f_{\mathrm{SB}}
  }
  \left(
    \Omega
    -
     2\pi \Delta f_{\mathrm{SB}}
  \right).
  \label{eq:mismatch_linearization}
\end{equation}
Assuming that the nonlinear phase and modal coupling vary negligibly
across the gain band, Eqs.~\eqref{eq:mismatch_linearization} and
\eqref{eq:gain_condition} provide the full gain bandwidth
\begin{equation}
   \Delta f_{\mathrm{BW}}(p_\mathrm{s},h)= \frac{2}{\pi}|\kappa(p_\mathrm{s})|L_{\mathrm{cav}} \Big/ \left| \frac{\mathrm{d}\phi_\mathrm{D}}{\mathrm{d}\Omega} \big|_{\Omega=2\pi \Delta f_{\mathrm{SB}}}\right|,
  \label{eq:bandwidth_general}
\end{equation}
where $\Delta f_{\mathrm{BW}}$ is the frequency
separation between the two edges of the gain band defined by $g(p_\mathrm{s},h,\Omega)=0$. It therefore represents the full gain bandwidth rather than the full width at half maximum (FWHM). Equation~\eqref{eq:bandwidth_general} is valid when the dispersion-phase slope is nonzero and varies little across the gain band. If these conditions are not satisfied, the two band edges must instead be found by solving $|\delta\phi(p_\mathrm{s},h,\Omega)| = 2|\kappa(p_\mathrm{s})|L_\mathrm{cav}$. An explicit fourth-order approximation for $\Delta f_{\mathrm{BW}}(p_\mathrm{s},h)$ is given by Eq.~\eqref{eq:bandwidth} in Appendix~\ref{app:small_signal_details}. 

Figure~\ref{fig:gaintheory}(a) shows the small-signal amplitude-growth rate for radial orders $p_\mathrm{s}=1,2$ and Floquet orders
$h=0,\ldots,5$. For both radial orders, the contribution of nonlinear phase $\Delta \phi_\mathrm{NL}$ is set to zero in this comparison. The two radial-mode families are vertically offset for clarity, while all 12 gain bands are normalized to the common reference $|\kappa(p_\mathrm{s}=1)|$. The band center is determined by the Floquet QPM condition $\delta\phi=0$. Within a given radial family, all Floquet orders have the same peak growth rate but different center frequencies and bandwidths. Because $|\kappa(2)|/|\kappa(1)|=3/4$, the $p_\mathrm{s}=2$ peaks reach
only $3/4$ of the $p_\mathrm{s}=1$ peak and have correspondingly
narrower gain bandwidths. Figure~\ref{fig:gaintheory}(b) compares two evaluations of $\Delta f_{\mathrm{BW}}$ as a function of
$\Delta f_{\mathrm{SB}}$ for $p_\mathrm{s}=1$ and $2$. The discrete markers show
$\Delta f_{\mathrm{BW}}=(\Omega_+-\Omega_-)/(2\pi)$, where
$\Omega_-$ and $\Omega_+$ are the two band-edge solutions of
$|\delta\phi(p_\mathrm{s},h,\Omega)|
=2|\kappa(p_\mathrm{s})|L_\mathrm{cav}$.
Here, $\delta\phi$ is evaluated using
Eq.~\eqref{eq:residual_mismatch} with the full Sellmeier-based
dispersive phase in Eq.~\eqref{eq:disp_phase}. The continuous curves
show the local approximation in Eq.~\eqref{eq:bandwidth_general}. The close agreement between the markers and curves indicates that the bandwidth narrowing at larger detuning is governed primarily by the increasing slope $|\mathrm{d}\phi_\mathrm{D}/\mathrm{d}\Omega|$.
\begin{figure}[h]
  \centering
  \includegraphics[width=0.48\textwidth]{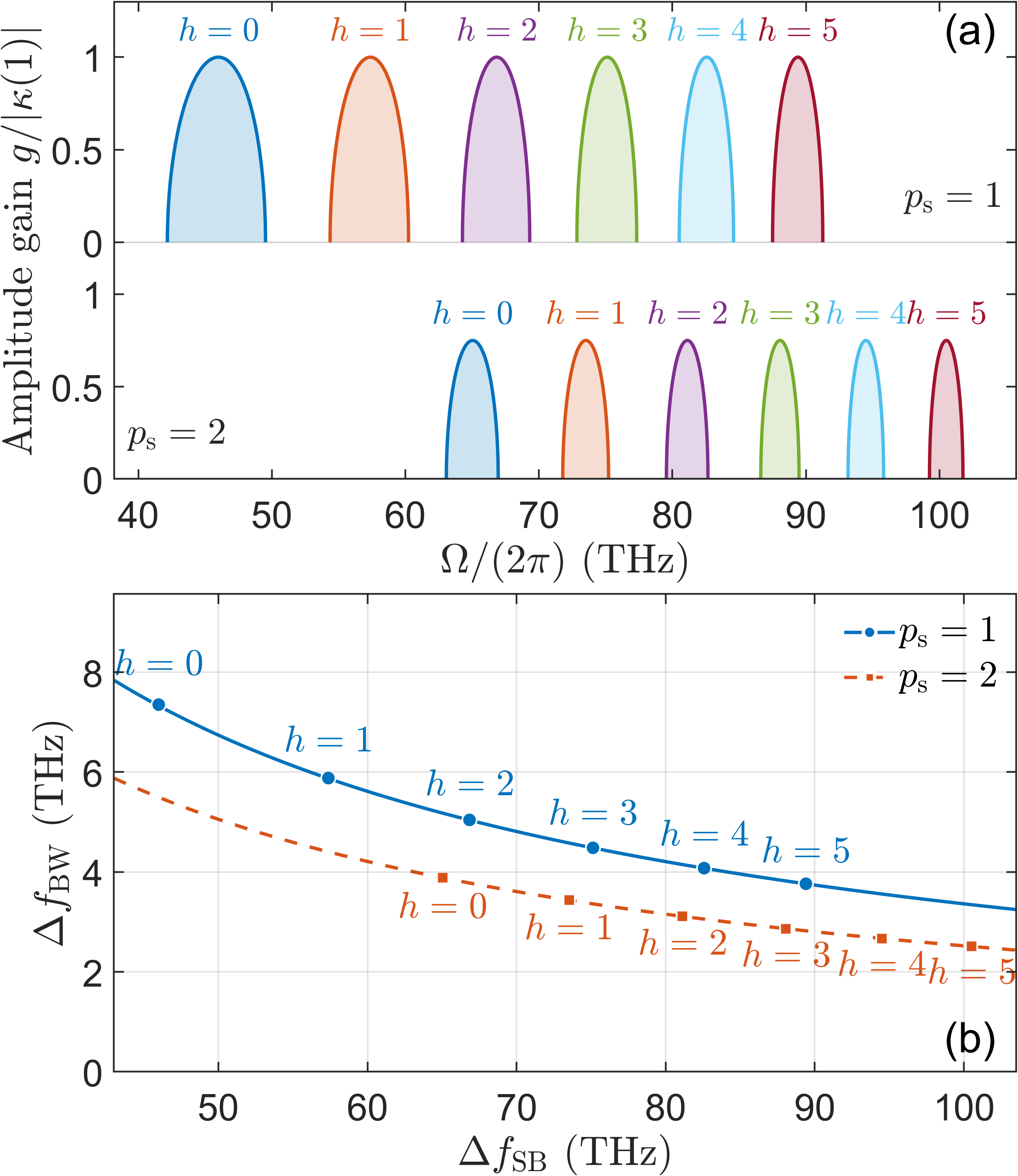}
\caption{Small-signal Floquet gain for a quasi-concentric MPC with
$C=1.95$ and $|\kappa(1)|L_\mathrm{cav}=0.9$.
(a) Normalized amplitude-growth rates for $p_\mathrm{s}=1,2$ and
$h=0,\ldots,5$. The two radial-mode families are vertically offset
for clarity. (b) Full gain bandwidth versus sideband detuning. Discrete markers are obtained from the two roots of the full Sellmeier-based gain-boundary equation, whereas continuous curves show the local approximation in Eq.~\eqref{eq:bandwidth_general}. Blue circles and the blue solid curve denote $p_\mathrm{s}=1$. Orange squares and the orange dashed curve denote $p_\mathrm{s}=2$.}
  \label{fig:gaintheory}
\end{figure}
 
\subsubsection{Large-Signal Regime}
\label{sec:largesignal}

\noindent The undepleted-pump approximation used in Sec.~\ref{subsec:small_signal} fails once a generated
signal--idler pair carries a non-negligible fraction of the input pump power. We therefore introduce a model containing the pump and one radial--Floquet pair $(p_\mathrm{s},h)$. This model assumes lossless propagation and equal initial signal and idler powers. The SPM--XPM contribution to the residual mismatch is evaluated using the input pump peak power $P_{\mathrm{p,in}}$ and is held constant during the subsequent pump-depleted evolution. Pump depletion is retained in the phase-sensitive FWM interaction, whereas depletion-induced changes of the nonlinear mismatch and coupling
between different sideband pairs are neglected. Appendix~\ref{app:pump_depleted_model} derives the model from the coupled-amplitude equations in Eq.~\eqref{eq:threewave}.

The residual mismatch evaluated at the input pump power is
\[
\delta\phi_{\mathrm{in}}(p_\mathrm{s},h,\Omega)
=
\phi_\mathrm{D}(\Omega)-\Delta N\Phi_0
+\Delta\phi_{\mathrm{NL,in}}(p_\mathrm{s})-2\pi h,
\]
where
$\Delta\phi_{\mathrm{NL,in}}(p_\mathrm{s})
=2[\gamma_\mathrm{SPM}-\gamma_\mathrm{XPM}(p_\mathrm{s})]
P_\mathrm{p,in}L_\mathrm{cav}$. Let
$\delta k_\mathrm{in}=\delta\phi_\mathrm{in}/L_\mathrm{cav}$
denote the residual phase mismatch per unit length. The transformed
signal and idler powers are denoted by $P_\mathrm{s}$ and
$P_\mathrm{i}$, respectively, and the converted fraction is defined as
$u=(P_\mathrm{s}+P_\mathrm{i})/P_\mathrm{p,in}$. After the variable substitution of $\kappa_{\mathrm{in}} = \left|\gamma_\mathrm{FWM}(p_\mathrm{s})\right| P_{\mathrm{p,in}}$, $z'=\kappa_{\mathrm{in}}z$, and $D_h = \delta k_{\mathrm{in}}/\kappa_{\mathrm{in}} = \delta\phi_{\mathrm{in}}/(\kappa_{\mathrm{in}}L_\mathrm{cav})$, the pump-depleted evolution for the growing separatrix selected by an
infinitesimal seed reduces to
\begin{equation}
  \left(
    \frac{\mathrm{d}u}{\mathrm{d}z'}
  \right)^2
  =
  4u^2(1-u)^2-D_h^2u^2.
  \label{eq:ueom}
\end{equation}
The right-hand side is non-negative only when $2(1-u)\geq|D_h|$. Consequently,
\begin{equation}
  u_{\max}
  =
  1-|D_h|/2,
  \qquad
  |D_h|\leq2,
  \label{eq:maximum_converted_fraction}
\end{equation}
and the instability disappears at $|D_h|=2$. At exact phase matching,
$D_h=0$, the lossless single-pair model permits $u_{\max}=1$. The
modal overlap determines $\kappa_\mathrm{in}$ and hence the conversion length, but not the ideal conversion limit. Linearizing
Eq.~\eqref{eq:ueom} for $u\ll1$ yields
$g=(\kappa_\mathrm{in}/2)\sqrt{4-D_h^2}$, recovering
Eq.~\eqref{eq:gain}. The finite- and zero-mismatch pump-depleted trajectories are given by Eqs.~\eqref{eq:separatrix_solution} and
\eqref{eq:separatrix_exact_matching}, respectively, in
Appendix~\ref{app:pump_depleted_model}.

Figure~\ref{fig:conversiontheory}(a) compares the pump-depleted
trajectories with their small-signal limits. The two models agree
during the initial growth but diverge once the generated sidebands
carry a significant fraction of the pump power. At $D_h=0$, the
pump-depleted solution approaches complete conversion, whereas the
small-signal solution grows without bound. At $D_h=1$, the converted
fraction reaches $u_{\max}=0.5$ and subsequently decreases through
coherent back-conversion. Figure~\ref{fig:conversiontheory}(b) summarizes the dependence of the maximum converted fraction on the normalized residual mismatch. At exact phase matching ($D_h=0$), the ideal model permits complete conversion, $u_{\max}=1$. As $|D_h|$ increases, $u_{\max}$ decreases linearly and reaches zero at the instability boundaries $D_h=\pm2$. Positive and negative mismatches give the same conversion limit because Eq.~\eqref{eq:ueom} depends on $D_h^2$. Exponential parametric growth therefore occurs only for $|D_h|<2$.
\begin{figure}[h]
  \centering
  \includegraphics[width=0.48\textwidth]{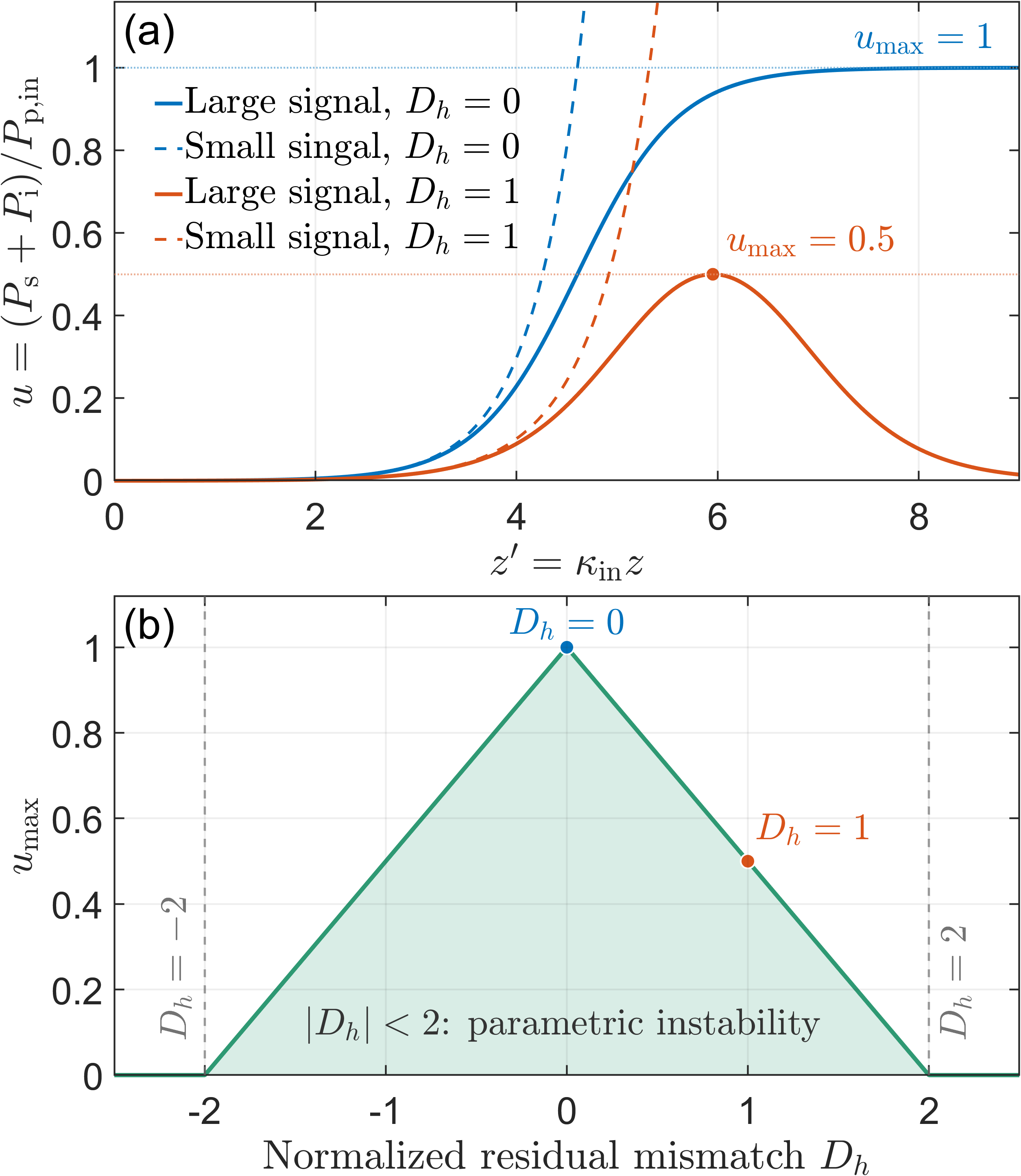}
  \caption{ Large-signal dynamics of the reduced pump--signal--idler model. (a) Converted fraction $u$ versus normalized
distance $z'$ for $D_h=0$ and $D_h=1$. Solid curves show the pump-depleted solutions, and dashed curves show their small-signal approximations. (b) Maximum converted fraction $u_{\max}=1-|D_h|/2$ versus normalized residual mismatch. Parametric instability exists for $|D_h|<2$.
}
  \label{fig:conversiontheory}
\end{figure}
\section{Numerical Model and Results}
\label{sec:numerics}

\noindent To test the above analytical predictions, we propagate the
multimode field through the MPC using a truncated MMGNLSE~\cite{Poletti2008,Wright2017-2}. This model retains the five
radially symmetric modes $\mathrm{LG}_{0p}$ with $p=0,\ldots,4$. With
$S_{pqrs}= S_{(0,p)(0,q)(0,r)(0,s)}$, the envelope of radial mode
$p$ obeys
\begin{equation}
  \begin{split}
  \frac{\partial A_p}{\partial z}
  &=
  -i\left[
    \beta(\omega_0+\Omega)
-\beta(\omega_0)
-\beta_1\Omega
  \right]A_p
  \\
  &\quad
  +i\frac{N_{0p}\Phi_0}{L_\mathrm{cav}}A_p
  +i\gamma_\mathrm{eff}(C)
  \sum_{q,r,s}S_{pqrs}A_qA_rA_s^{*}.
  \end{split}
  \label{eq:mmgnlse}
\end{equation}
The first term describes material dispersion in the pump group-velocity frame, using the full argon Sellmeier relation
\cite{Dalgarno1960,Peck1964,Borzsonyi2008}. The second gives the
mode-dependent Gouy phase. Both terms are diagonal in radial-mode
index. The nonlinear sum is restricted to the pump self-overlap
$S_{0,0,0,0}$ and the pump-mediated diagonal overlaps
$S_{p,0,0,p}$. The model thereby retains pump SPM, pump-induced XPM,
phase-sensitive FWM, and, in the large-signal calculations,
common-pump back-action. Direct nonlinear mixing between two
different higher-order radial indices is omitted. The corresponding equations, input profile, and complete simulation
parameters are given in Appendix~\ref{app:numerical_implementation}. The nonlinear refractive index is scaled linearly with gas density from $n_2=10.3\times10^{-20}$~cm$^2$/W at $1$~bar \cite{Bree2010,Zahedpour2015}. 

The input is launched in the fundamental mode, while the higher-order modes are seeded with independent complex Gaussian noise corresponding to one photon per spectral mode~\cite{Dudley2002,Frosz2010}. Because only diagonal radial FWM channels with $p_\mathrm{i}=p_\mathrm{s}$ are retained, each higher-order modal envelope contains both the signal and idler components. We denote the pump-mode energy by $\mathcal{E}_{\mathrm{p}}(J)$ and the combined signal--idler energy in radial mode $p_\mathrm{s}$ by $\mathcal{E}_{\mathrm{s+i}}(p_\mathrm{s},J)$. The latter is integrated over the full modal spectrum and therefore includes both symmetric sidebands and all retained Floquet orders. The total energy is defined as $\mathcal{E}_{\mathrm{tot}}(J)
=\mathcal{E}_{\mathrm{p}}(J)
+\sum_{p_\mathrm{s}=1}^{4}
\mathcal{E}_{\mathrm{s+i}}(p_\mathrm{s},J)$,
with $E_{\mathrm{in}}=\mathcal{E}_{\mathrm{p}}(0)$. For a specific Floquet branch,
$\mathcal{E}_{\mathrm{s+i}}(p_\mathrm{s},h,J)$ denotes the combined
energy integrated over the symmetric signal and idler spectral windows associated with $(p_\mathrm{s},h)$.

\subsection{Small-signal comparison}

\begin{figure}[h]
  \centering
  \includegraphics[width=0.48\textwidth]{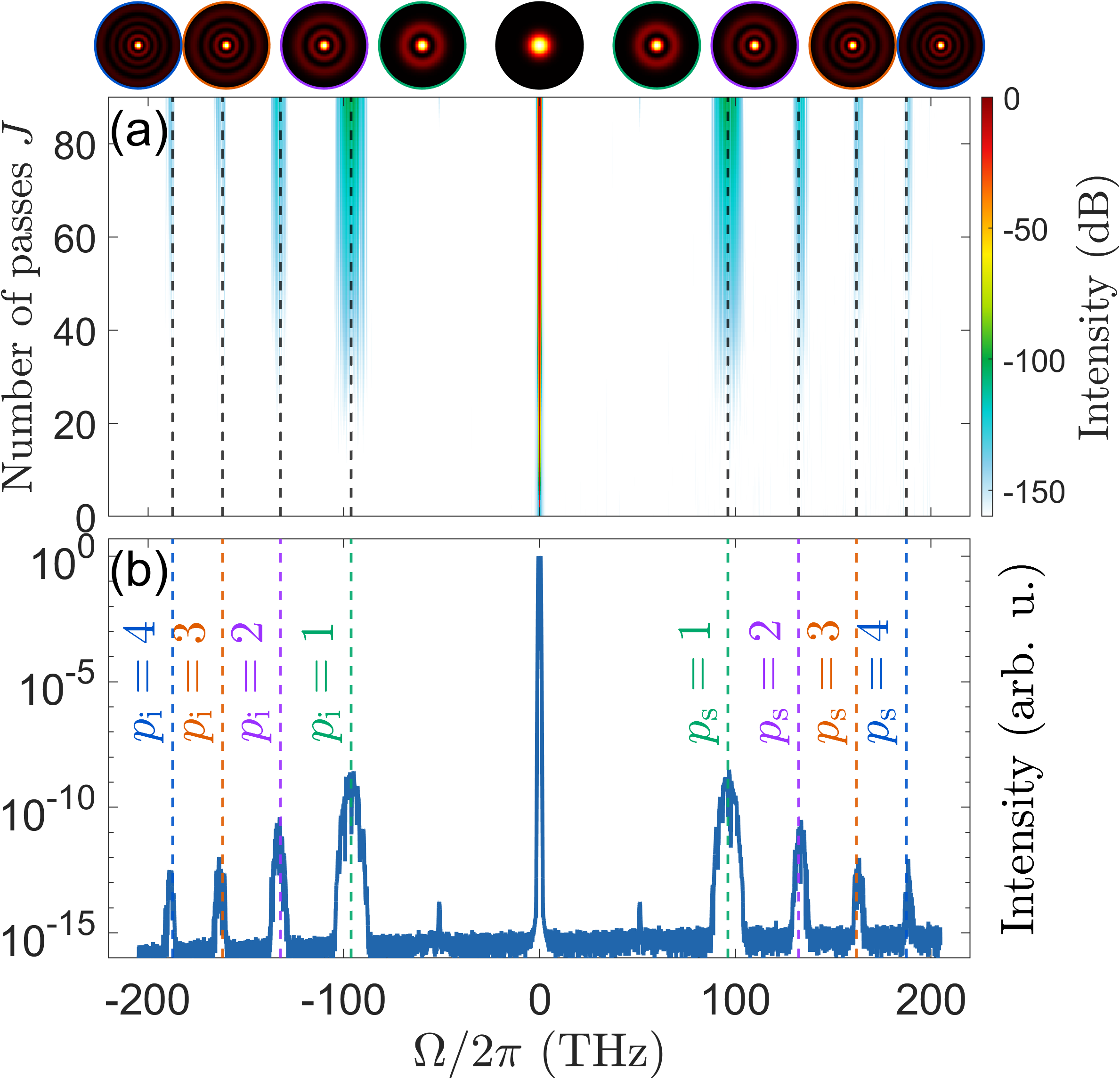}
 \caption{Small-signal GPI in a quasi-collimated MPC ($C=0.05$). (a) Spectral intensity versus pass number $J$. Dashed lines mark the Floquet-QPM sideband frequencies, and circular insets
show the corresponding $\mathrm{LG}_{0p}$ intensity profiles.
(b) Spectra after $J=90$ passes, with colored vertical lines marking
the predicted signal and idler detunings by Floquet theory.}
  \label{fig:smallsignal}
\end{figure}

\noindent
Figure~\ref{fig:smallsignal}(a) shows the spectral
evolution in the quasi-collimated geometry under different numbers of passes $J$. Starting from the semiclassical stochastic seed, four symmetric pairs of sidebands associated with the radial orders $p_\mathrm{s}=1,\ldots,4$ emerge on both sides of the pump. The amplified bands remain centered at nearly constant detunings as they grow, indicating that their frequencies are set primarily by the Floquet phase-matching condition. The dashed vertical lines show the predicted $h=0$ sideband frequencies, and their agreement with the simulated bands supports the Floquet QPM model. The mode profiles above the map identify the higher-order radial modes associated with the corresponding sidebands. The pump near $\Omega=0$ remains dominant throughout the propagation, consistent with the small-signal regime.

Figure~\ref{fig:smallsignal}(b) shows the spectrum after $J=90$
passes. The four signal--idler pairs appear as distinct spectral peaks on both sides of the pump. Their peak positions agree with the corresponding Floquet-QPM predictions on both sides of the pump. The $p_\mathrm{s}=1$ pair is the strongest, while the higher radial orders are progressively weaker because their modal overlaps and FWM coupling coefficients are smaller. The symmetric locations of the signal and idler peaks about the pump are consistent with the degenerate FWM frequencies $\omega_0\pm\Omega$.

\begin{figure}[h]
  \centering
  \includegraphics[width=0.48\textwidth]{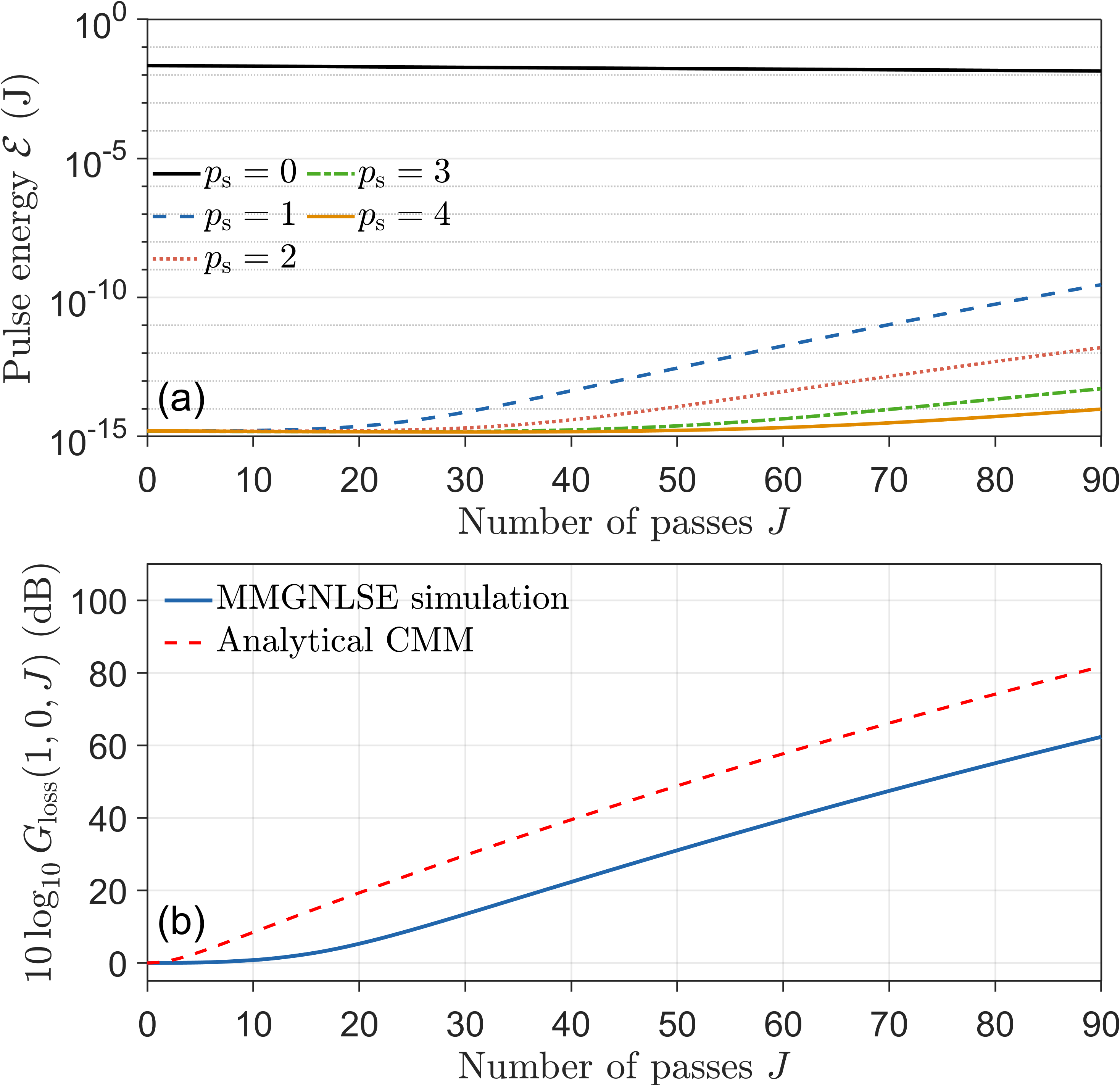}
  \caption{ Small-signal gain comparison for $C=0.05$.
(a) Pump-mode energy $\mathcal{E}_{\mathrm{p}}(J)$ and combined
signal--idler energies $\mathcal{E}_{\mathrm{s+i}}(p_\mathrm{s},J)$ for $p_\mathrm{s}=1,\ldots,4$.
(b) Noise-seeded relative sideband-energy gain $G_{\mathrm{loss}}(p_\mathrm{s},h,J)$ for the $p_\mathrm{s}=1$, $h=0$ sideband from the MMGNLSE simulation (blue) and analytical CMM (red dashed).
}
  \label{fig:smallsignalgain}
\end{figure}

Figure~\ref{fig:smallsignalgain}(a) shows the pump-mode pulse energy
$\mathcal{E}_{\mathrm{p}}(J)$ for $p=0$ and the combined
signal--idler energies $\mathcal{E}_{\mathrm{s+i}}(p_\mathrm{s},J)$ for $p_\mathrm{s}=1,\ldots,4$. These energies are obtained by integrating the spectral intensity resulting from the numerical solution of MMGNLSE accounting for the mirror loss. The $p_\mathrm{s}=1$ mode grows fastest because it has the largest overlap and hence the largest coupling rate $|\kappa(p_\mathrm{s})|$. The progressively slower growth of the higher radial orders follows the corresponding decrease in modal overlap. The noise-seeded gain is obtained from the numerical MMGNLSE solution. We define $G_\mathrm{loss} (p_\mathrm{s}, h, J)$ as the ratio between the output and input energies after $J$ passes. The mirror loss is considered through the effective pass number $J_{\mathrm{eff}}(J) =(1-\mathcal{R}^{J})/(1-\mathcal{R})$ with accumulated transmission $\mathcal{R}^{J}$. Applying these corrections to Eq.~\eqref{eq:noise_seeded_gain_exact} gives the following loss-modified gain:
\begin{equation}
  G_{\mathrm{loss}}(p_\mathrm{s},h,J) = \mathcal{R}^{J}
  \cosh\!\left[
    2|\kappa_{\mathrm{in}}|L_\mathrm{cav}
    J_\mathrm{eff}(J)
  \right].
  \label{eq:lossy_noise_seeded_gain}
\end{equation}
For the MMGNLSE simulation, the cumulative noise-seeded relative
energy gain is defined as $G_{\mathrm{loss}}(p_\mathrm{s},h,J) =\mathcal{E}_{\mathrm{s+i}}(p_\mathrm{s},h,J)/
\mathcal{E}_{\mathrm{s+i}}(p_\mathrm{s},h,0)$.
The branch energy is calculated by integrating the simulated spectral
intensity over the symmetric signal and idler windows satisfying
$\left||\Omega|/(2\pi)-\Delta f_{\mathrm{SB}}(p_\mathrm{s},h)\right|<1~\mathrm{THz}$.

Figure~\ref{fig:smallsignalgain}(b) compares the gain of the
$p_\mathrm{s}=1$, $h=0$ branch obtained from the MMGNLSE simulation
with the analytical CMM prediction in
Eq.~\eqref{eq:lossy_noise_seeded_gain}. The two curves in Fig.~\ref{fig:smallsignalgain}(b) exhibit the same
overall growth trend. At $J=90$, the simulation gives approximately $63$~dB, about $18$~dB below the analytical CMM prediction of
approximately $81$~dB when the values on the decibel scale are compared. The analytical CMM assumes exact phase matching and applies the input peak power to the entire pulse. The MMGNLSE simulation instead integrates the gain over the Gaussian temporal profile and a finite spectral window, leading to slower pulse-integrated growth.

\subsection{Large-signal dynamics and multi-order GPI}

\begin{figure}[t]
  \centering
  \includegraphics[width=0.48\textwidth]{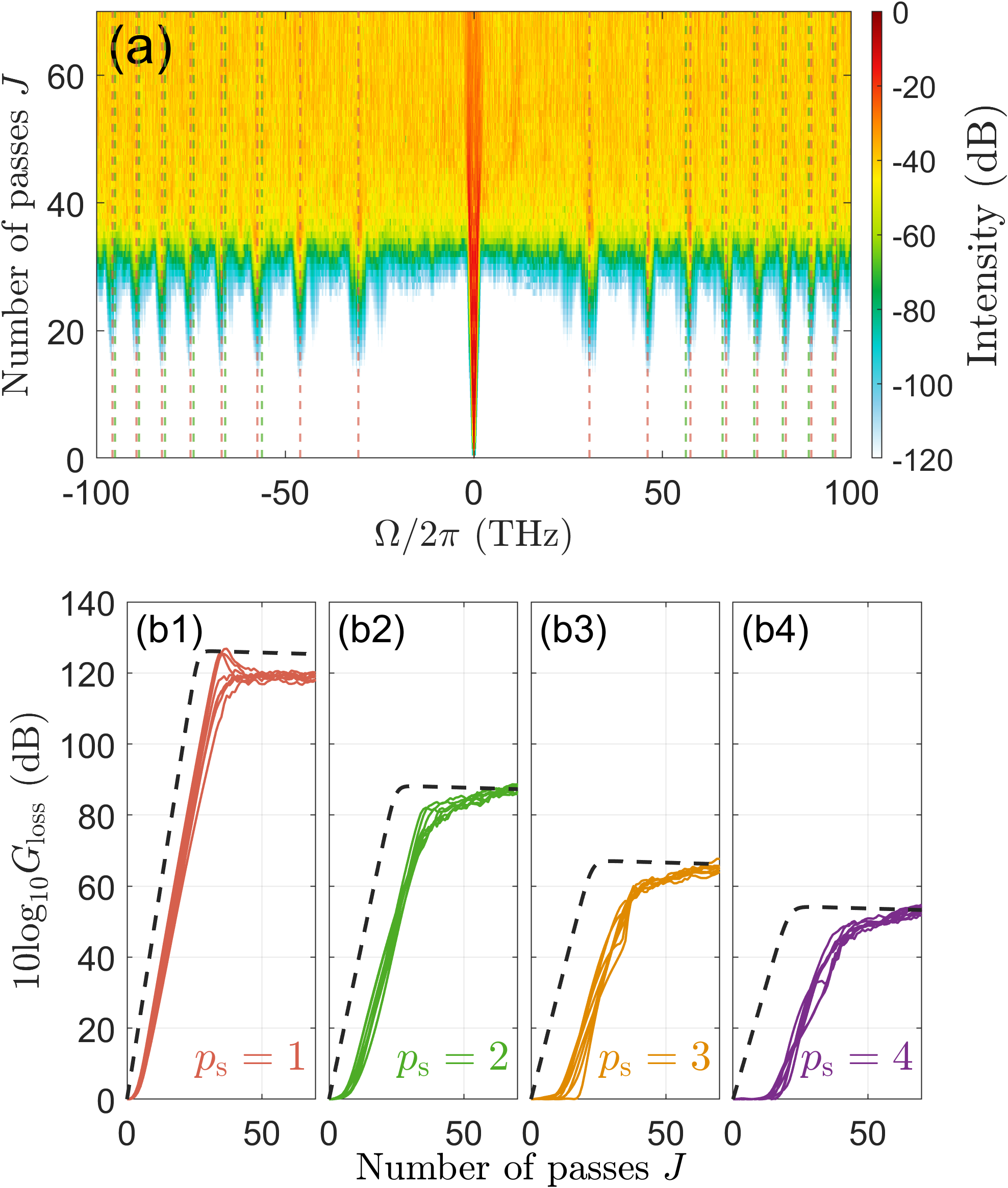}
  \caption{Large-signal multichannel GPI in a quasi-concentric MPC
($C=1.95$). (a) Spectral evolution summed over $p=0,1,2$. Dashed
lines mark the Floquet-QPM frequencies. (b1)--(b4) Noise-seeded
relative gain for $p_\mathrm{s}=1,\ldots,4$. Colored curves denote
different Floquet orders, and black dashed curves show qualitative
predictions of the isolated pump-depleted CMM.}
  \label{fig:largesignal}
\end{figure}

\noindent
For $C=1.95$, the large single-pass Gouy phase allows the Floquet QPM
condition to be satisfied in normal dispersion for multiple radial and Floquet orders. We use the branch-resolved gain
$G_{\mathrm{loss}}(p_\mathrm{s},h,J)$ defined above. Figure~\ref{fig:largesignal}(a) shows the spectral evolution summed over the pump mode $p=0$ and the two higher-order radial modes $p_\mathrm{s}=1,2$. The dashed vertical lines mark the sideband frequencies obtained by solving the Floquet QPM condition in Eq.~\eqref{eq:Floquet_QPM}. The simulated peaks remain close to these predicted frequencies during both initial growth and the subsequent pump-depletion regime. The sideband positions are therefore governed primarily by the imbalance of material-dispersion phase and the single-pass Gouy phase, with only a modest shift arising from the evolving nonlinear phase. 

Figures~\ref{fig:largesignal}(b1)--\ref{fig:largesignal}(b4) show the noise-seeded gain of the radial modes $p_\mathrm{s}=1,\ldots,4$. The colored curves within each panel represent different Floquet orders $h$, rather than independent noise realizations. During the initial stage, their growth rates follow the modal overlaps. The $p_\mathrm{s}=1$ channel grows fastest, whereas the higher radial orders remain progressively weaker. As the dominant $p_\mathrm{s}=1$ channel extracts an appreciable fraction of the pump energy, its evolution departs from the undepleted-pump approximation. The higher-order channels show a clear gain saturation within the simulated range, but their growth is still modified by depletion of the common pump and should not be interpreted as independent single-channel saturation. The black dashed curves are qualitative guides from the pump-depleted CMM, calculated using the same path-averaged coupling coefficients. The $p_\mathrm{s}=1$ curves show a modest overshoot near $J\simeq30$. This behavior may involve coherent back-conversion. 

\begin{figure}[tb]
  \centering
  \includegraphics[width=0.48\textwidth]{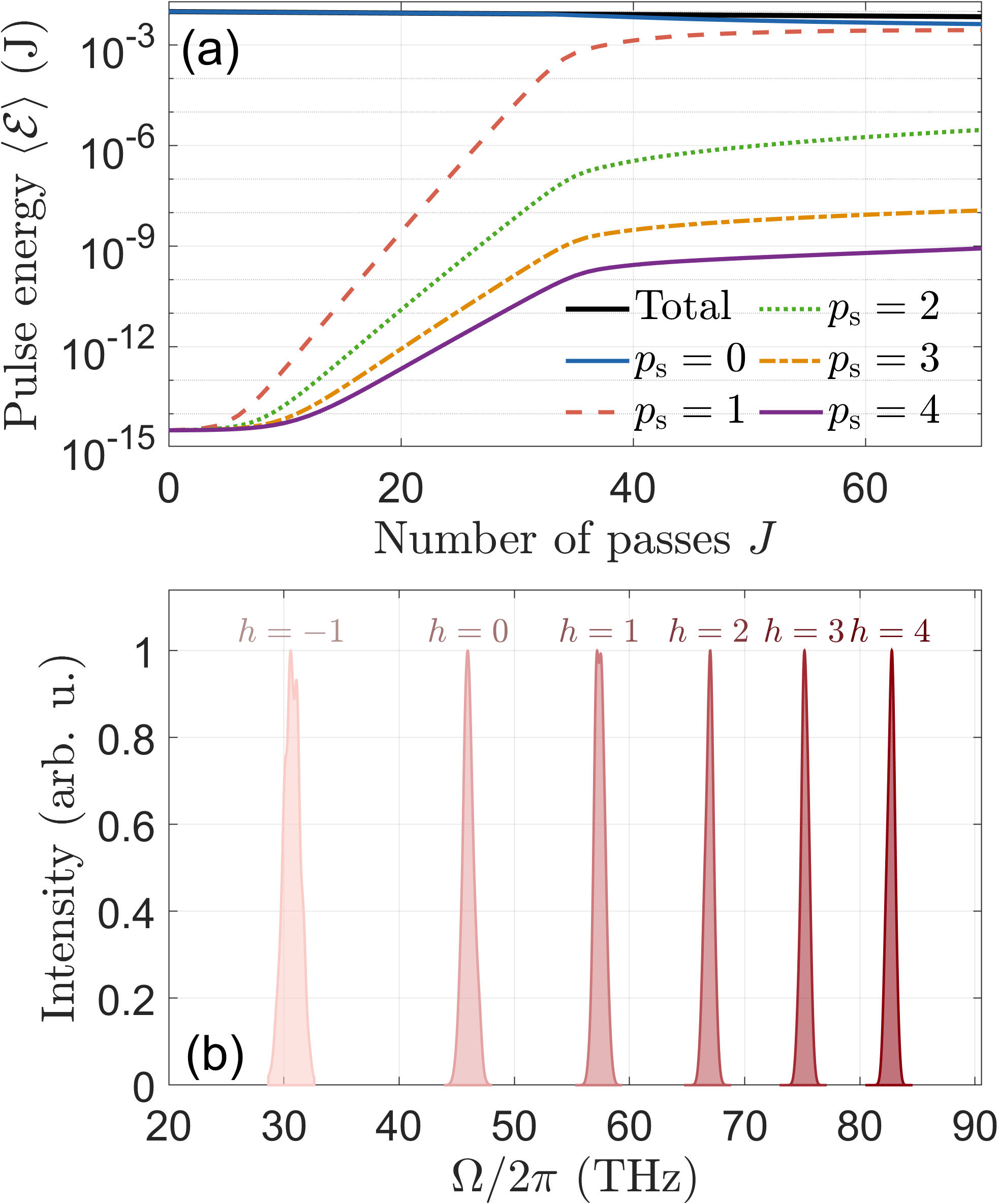}
 \caption{Shared-pump dynamics for $C=1.95$.
(a) Ensemble-averaged pump energy
$\mathcal{E}_{\mathrm{p}}(J)$ and combined signal--idler energies
$\mathcal{E}_{\mathrm{s+i}}(p_\mathrm{s},J)$ for
$p_\mathrm{s}=1,\ldots,4$. The black curve is
$\mathcal{E}_{\mathrm{tot}}(J)$. (b) Ensemble-averaged $p_\mathrm{s}=1$ spectrum at
$J=20$, with each $h=-1,\ldots,4$ band normalized independently.}
  \label{fig:modedynamics}
\end{figure}

Figure~\ref{fig:modedynamics}(a) shows the ensemble-averaged pump
energy $\langle\mathcal{E}_{\mathrm{p}}(J)\rangle$ and combined
signal--idler energies $\langle\mathcal{E}_{\mathrm{s+i}}(p_\mathrm{s},J)\rangle$ over $100$ independent noise realizations. As the GPI sidebands develop, the energy in the fundamental pump mode decreases while the higher-order radial modes gain energy, with the $p_\mathrm{s}=1$ mode receiving the largest energy. This correlated evolution supports pump-to-sideband energy transfer through the retained nonlinear channels. The higher-order-mode energies approach lower plateaus, reflecting shared-pump depletion rather than independent single-channel saturation. Figure~\ref{fig:modedynamics}(b) shows the ensemble-averaged spectrum of the $p_\mathrm{s}=1$ mode at $J=20$, before strong pump depletion. Each Floquet band with $h=-1,\ldots,4$ is normalized to its own maximum, so the panel compares spectral shapes rather than absolute gains. The bands become progressively narrower at larger detuning because the increasing slope $|\mathrm{d}\phi_\mathrm{D}/\mathrm{d}\Omega|$ maps the allowed residual-phase interval onto a smaller frequency interval. This trend is consistent with Eq.~\eqref{eq:bandwidth_general}.

\subsection{Geometric and pressure tunability}

\begin{figure}[t]
  \centering
  \includegraphics[width=0.46\textwidth]{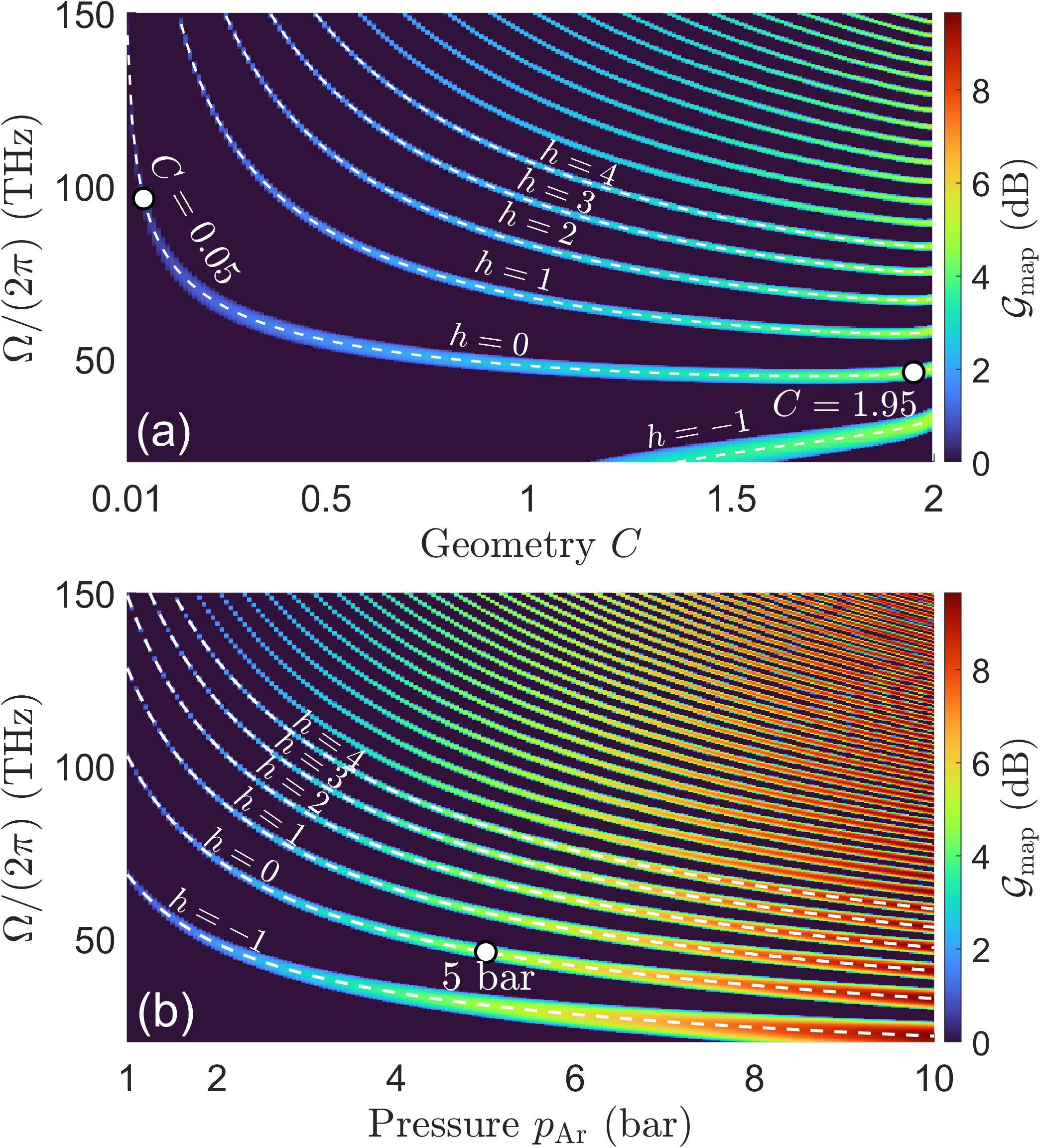}
  \caption{Tunability of the $p_\mathrm{s}=1$ GPI sidebands. The color scale shows $(20/\ln 10)G$, and white dashed curves mark the
phase-matched centers for $h=-1,\ldots,4$. (a) Gain versus geometry parameter $C$ and detuning $\Omega/(2\pi)$ at $5$~bar. The circles mark $C=0.05$ and $1.95$. (b) Gain versus argon pressure and detuning $\Omega/(2\pi)$ at $C=1.95$. The circle marks $5$~bar.}
  \label{fig:tunability}
\end{figure}

\noindent A remarkable feature of GPI in an MPC is that the sideband detuning can be controlled through either the cavity geometry or the gas pressure. For fixed $p_\mathrm{s}=1$ and varied Floquet order $h$, we express the small-signal growth accumulated during
one pass as
\begin{equation}
  G(h,\Omega;C,p_\mathrm{Ar})
  =
  \!\left[
    g(h,\Omega;C,p_\mathrm{Ar})
  \right]
  L_\mathrm{cav}(C),
  \label{eq:gain_per_pass}
\end{equation}
where $g$ is given by Eq.~\eqref{eq:gain} and $L_\mathrm{cav}(C)=CR_0$. For fixed $p_\mathrm{s}=1$, the remaining four factors, $h$, $\Omega$, $C$ and $p_\mathrm{Ar}$, affect the gain $G$. At $R_0=1$~m, varying $C$ changes both the Floquet phase-matching condition and the nonlinear coupling. As Eq.~\eqref{eq:gain_per_pass} is the amplitude gain, the intensity gain in decibels shown in Fig.~\ref{fig:tunability} corresponds to $10\log_{10} \mathrm{e}^{2G}$. The white dashed curves are the calculated phase-matched band centers obtained by solving $\delta\phi (p_\mathrm{s}=1,h,\Omega)=0$ for the Floquet orders $h=-1,\ldots,4$. The sideband positions in Fig.~\ref{fig:tunability}(a) therefore reflect the combined variation of the Gouy and material-dispersion phases. The color scale additionally depends on the geometry-dependent coefficient $\gamma_\mathrm{eff}(C)=\gamma_0(C)F(C)$. At a fixed pressure of $5$~bar, the sideband gain is calculated for $h$ from -1 to 17. Varying $C$ from $0.05$ to $1.95$ shifts the $h=0$ sideband detuning from approximately $96$ to $46$~THz. The additional low-detuning band at large $C$ corresponds to the $h=-1$ Floquet branch. For $p_\mathrm{s}=1$,
$\Delta\phi_\mathrm{NL}(1)=0$, and the phase-matching condition becomes $\phi_\mathrm{D}(\Omega)=\Delta N\Phi_0-2\pi$. Over the detuning range considered here, argon is normally dispersive,
so $\phi_\mathrm{D}(\Omega)$ is positive and increases with
$|\Omega|$. A nonzero-detuning solution therefore exists only when
$\Delta N\Phi_0>2\pi$, which occurs at sufficiently large $C$.

Gas pressure provides a complementary tuning parameter because both the argon refractivity and its Kerr coefficient scale approximately with gas density. Increasing the argon pressure strengthens the material dispersion, so $\phi_\mathrm{D}(\Omega)$ increases at a fixed detuning. This is the dominant reason why the phase-matched sidebands shift toward smaller $|\Omega|$ in Fig.~\ref{fig:tunability}(b), although the nonlinear
phase contribution also varies with pressure. At the same time, the
increase in $n_2$ strengthens the nonlinear coupling and hence the
gain. Gain spectra are calculated for $h=-1,\ldots,36$, with the
phase-matched band centers for $h=-1,\ldots,4$ marked by white
dashed curves. At fixed $C=1.95$, the $h=0$ sideband detuning consequently shifts from approximately $103$ to $33$~THz as the pressure increases from $1$ to $10$~bar [Fig.~\ref{fig:tunability}(b)]. For $p_\mathrm{s}=1$ where $\Delta\phi_\mathrm{NL}(1)=0$, the frequency shift is governed primarily by the pressure dependence of the material dispersion, whereas the gain variation also reflects the pressure-dependent Kerr coefficient.

For comparison, the first experimental observation of GPI in a
standard GRIN multimode fiber pumped at $1064$~nm reported a
first-order sideband detuning of $123.5$~THz, in close agreement with
the theoretical prediction of approximately $125$~THz
\cite{Krupa2016}. Under the present MPC conditions at $1030$~nm, the
$p_\mathrm{s}=1$, $h=0$ detuning instead varies from approximately
$96$ to $46$~THz as the cavity geometry is tuned. The comparison highlights the geometry-dependent tunability of the MPC sidebands.

%

\section{Conclusion}
\label{sec:conclusion}
 
\noindent
We have developed a Floquet framework for GPI in nonlinear MPCs. Unlike GPI in GRIN fibers, where quasi-phase matching arises from a continuous self-imaging modulation~\cite{Krupa2016}, the MPC mechanism is governed by the discrete modal Gouy-phase imbalance accumulated during each mirror-to-mirror pass. It can therefore be driven by a single-mode Gaussian pump. The resulting sidebands are tunable through the cavity geometry and gas pressure. For example, the detuning of the $p_\mathrm{s}=1$, $h=0$ branch in the quasi-concentric MPC is approximately $46$~THz. The analytical model describes both the small-signal gain and the pump-depleted evolution of an isolated pump--signal--idler channel. The CMM predicts that the maximum converted fraction decreases linearly with the magnitude of the normalized residual mismatch. Truncated MMGNLSE simulations reproduce the predicted Floquet sideband positions and show depletion of the common pump by the dominant $p_\mathrm{s}=1$ channel and competition among the higher-order channels. Direct energy transfer between different higher-order radial modes is not included in the present model.

GPI may impose a limitation on spatial beam quality, while also providing a tunable mechanism for broadband multicolor generation. In practical MPCs, finite apertures and mode-selective losses can preferentially suppress higher-order modes~\cite{Guo2024}. Because the simulations include only uniform mirror loss, their predicted growth should be interpreted as an idealized limit rather than a quantitative experimental threshold. Finally, the modal selection rules allow counter-rotating vortex sidebands from a Gaussian pump, suggesting a route to multicolor light carrying orbital angular momentum.

\appendix

\section{Derivation of the pump-depleted CMM}
\label{app:pump_depleted_model}

The reduced large-signal model used in Sec.~\ref{sec:largesignal}
contains the pump and one signal--idler pair and retains only their
phase-sensitive FWM interaction. After the constant SPM and XPM phase
rotations evaluated at $P_{\mathrm{p,in}}$ have been transformed out,
the amplitudes obey~\cite{Boyd2020,Armstrong1962}
\begin{align}
  \frac{\mathrm{d}B_\mathrm{p}}{\mathrm{d}z}
  &=
  i\,2\gamma_\mathrm{FWM}(p_\mathrm{s})
  B_\mathrm{p}^{*}B_\mathrm{s}B_\mathrm{i}
  \exp(i\delta k_\mathrm{in}z),
  \nonumber
  \\
  \frac{\mathrm{d}B_\mathrm{s}}{\mathrm{d}z}
  &=
  i\,\gamma_\mathrm{FWM}(p_\mathrm{s})
  B_\mathrm{p}^{2}B_\mathrm{i}^{*}
  \exp(-i\delta k_\mathrm{in}z),
  \label{eq:threewave}
  \\
  \frac{\mathrm{d}B_\mathrm{i}}{\mathrm{d}z}
  &=
  i\,\gamma_\mathrm{FWM}(p_\mathrm{s})
  B_\mathrm{p}^{2}B_\mathrm{s}^{*}
  \exp(-i\delta k_\mathrm{in}z).
  \nonumber
\end{align}
Here $\delta k_\mathrm{in}
=\delta\phi_\mathrm{in}(p_\mathrm{s},h,\Omega)/L_\mathrm{cav}$.
The factor of two in the pump equation accounts for the two pump
photons involved in each degenerate FWM event.
Equations~\eqref{eq:threewave} conserve
$P_\mathrm{p}+P_\mathrm{s}+P_\mathrm{i}=P_\mathrm{p,in}$, where
$P_j=|B_j|^2$ are envelope powers normalized to the common reference
frequency $\omega_0$. For symmetric signal and idler seeds,
$P_\mathrm{s}=P_\mathrm{i}$. Defining
$u = (P_\mathrm{s}+P_\mathrm{i})/P_\mathrm{p,in}
=1-P_\mathrm{p}/P_\mathrm{p,in}$ from power conservation and writing
$B_j=\sqrt{P_j}\exp(i\theta_j)$, we introduce the relative phase
\begin{equation}
  \varphi
  =
  \theta_\mathrm{s}
  +
  \theta_\mathrm{i}
  -
  2\theta_\mathrm{p}
  +
  \delta k_\mathrm{in}z.
  \label{eq:relative_phase}
\end{equation}
The coupled-amplitude equations then reduce to
\begin{align}
  \frac{\mathrm{d}u}{\mathrm{d}z'}
  &=
  2u(1-u)\sin\varphi,
  \nonumber
  \\
  \frac{\mathrm{d}\varphi}{\mathrm{d}z'}
  &=
  D_h+2(1-2u)\cos\varphi,
  \label{eq:power_phase_system}
\end{align}
where
$\kappa_\mathrm{in}
=|\gamma_\mathrm{FWM}(p_\mathrm{s})|P_\mathrm{p,in}$,
$z'=\kappa_\mathrm{in}z$, and
$D_h=\delta k_\mathrm{in}/\kappa_\mathrm{in}$. They possess the conserved Hamiltonian
\begin{equation}
  \mathcal{H}
  =
  2u(1-u)\cos\varphi+D_hu.
  \label{eq:large_signal_hamiltonian}
\end{equation}
The growing trajectory selected by the limit $u(0)\rightarrow0^+$ is
the separatrix with $\mathcal{H}=0$. Hence
$2(1-u)\cos\varphi=-D_h$. Eliminating $\varphi$ from
Eq.~\eqref{eq:power_phase_system} then yields Eq.~\eqref{eq:ueom} and
the turning point in Eq.~\eqref{eq:maximum_converted_fraction}.

For $0<|D_h|<2$, the corresponding separatrix solution is
\begin{equation}
  u(z')
  =
  \frac{1-a_h^2}
       {1+a_h\cosh\!\left[
          2\sqrt{1-a_h^2}\,(z'-z'_\mathrm{t})
        \right]},
  \label{eq:separatrix_solution}
\end{equation}
where $a_h=|D_h|/2$ and $z'_\mathrm{t}$ is the seed-dependent
position of the turning point. The converted fraction reaches
$u(z'_\mathrm{t})=1-|D_h|/2$ and then decreases through coherent
back-conversion. At exact phase matching, the growing solution becomes
\begin{equation}
  u(z')
  =
  \frac{1}{1+\exp[-2(z'-z'_0)]},
  \qquad D_h=0,
  \label{eq:separatrix_exact_matching}
\end{equation}
where $z'_0$ is the normalized distance at which $u=1/2$ and is
fixed by the initial converted fraction $u(0)$. When $D_h=0$, $u$ approaches unity asymptotically and no turning point occurs.

This closed-form reduction keeps the nonlinear mismatch fixed at its
input-power value. It therefore does not include depletion-induced
changes of pump SPM and sideband XPM, nonlinear phase shifts generated by strongly populated sidebands, linear loss, or competition among several radial and Floquet channels. Those effects are included only to the extent specified by the truncated multimode simulations in Sec.~\ref{sec:numerics}.

\section{Small-signal coupled-mode solution and bandwidth approximation}
\label{app:small_signal_details}

Let $\delta k =\delta\phi/L_\mathrm{cav}$. Under the undepleted-pump approximation, $\kappa(p_\mathrm{s})
=\gamma_\mathrm{FWM}(p_\mathrm{s})P_\mathrm{p}$ is constant.
Suppressing the arguments for compactness, we remove the explicit
mismatch phase by defining
$\widetilde{A}_\mathrm{s}
=A_\mathrm{s}\exp(i\delta k z/2)$ and
$\widetilde{A}_\mathrm{i}^{*}
=A_\mathrm{i}^{*}\exp(-i\delta k z/2)$.
The linearized signal--idler equations then become
\begin{equation}
  \frac{\mathrm{d}}{\mathrm{d}z}
  \begin{pmatrix}
    \widetilde{A}_\mathrm{s}\\
    \widetilde{A}_\mathrm{i}^{*}
  \end{pmatrix}
  =
  \begin{pmatrix}
    i\delta k/2 & i\kappa\\
    -i\kappa^{*} & -i\delta k/2
  \end{pmatrix}
  \begin{pmatrix}
    \widetilde{A}_\mathrm{s}\\
    \widetilde{A}_\mathrm{i}^{*}
  \end{pmatrix}.
  \label{eq:small_signal_matrix}
\end{equation}
The characteristic equation of this matrix yields
\begin{equation}
  g^2
  =
  |\kappa|^2
  -
  \left(\delta k/2\right)^2.
  \label{eq:gain_eigenvalue}
\end{equation}
Restoring the arguments and using
$\delta k(p_\mathrm{s},h,\Omega)
=\delta\phi(p_\mathrm{s},h,\Omega)/L_\mathrm{cav}$ then yields
Eq.~\eqref{eq:gain} in the main text.

For statistically independent signal and idler seeds, the
phase-sensitive cross terms vanish upon ensemble averaging. Within a
positive-gain band, the relative total sideband power gain is
\begin{equation}
  \begin{split}
  \mathcal{G}(p_\mathrm{s},h,\Omega,J)
  &=
  \frac{
    \left\langle
      |A_\mathrm{s}(Z_J)|^2+|A_\mathrm{i}(Z_J)|^2
    \right\rangle
  }{
    \left\langle
      |A_\mathrm{s}(0)|^2+|A_\mathrm{i}(0)|^2
    \right\rangle
  }
  \\
  &=
  1+
  \frac{2|\kappa(p_\mathrm{s})|^2}
       {g^2(p_\mathrm{s},h,\Omega)}
  \sinh^2\!\left[
    g(p_\mathrm{s},h,\Omega)Z_J
  \right].
  \end{split}
  \label{eq:noise_seeded_power_gain}
\end{equation}
At exact Floquet phase matching, $g(p_\mathrm{s},h,\Omega)=|\kappa(p_\mathrm{s})|$, and
Eq.~\eqref{eq:noise_seeded_power_gain} reduces to
Eq.~\eqref{eq:noise_seeded_gain_exact}.

The local bandwidth formula in Eq.~\eqref{eq:bandwidth_general} can
be written explicitly when the dispersive phase is retained through
fourth order,
$\phi_\mathrm{D}(\Omega)\simeq
L_\mathrm{cav}(\beta_2\Omega^2+\beta_4\Omega^4/12)$. This gives
\begin{equation}
  \begin{split}
  \Delta f_\mathrm{BW}(p_\mathrm{s},h)
  &\simeq
  \frac{|\kappa(p_\mathrm{s})|}
  {\pi\left|
    \beta_2\Omega_\mathrm{SB}(p_\mathrm{s},h)
    +\dfrac{\beta_4}{6}
    \Omega_\mathrm{SB}^3(p_\mathrm{s},h)
  \right|},
  \\
  \Omega_\mathrm{SB}(p_\mathrm{s},h)
  &=
  2\pi\Delta f_\mathrm{SB}(p_\mathrm{s},h).
  \end{split}
  \label{eq:bandwidth}
\end{equation}
This approximation requires the dispersion-phase slope to remain
nonzero and approximately constant across the band. Otherwise, the full gain bandwidth must instead be obtained from the two solutions of $\left| \delta\phi(p_\mathrm{s},h,\Omega)\right| = 2|\kappa(p_\mathrm{s})|L_\mathrm{cav}$.

\section{Truncated multimode equations and numerical implementation}
\label{app:numerical_implementation}

\begin{table}[t]
\caption{Complete simulation parameters for the quasi-collimated
small-signal case and the quasi-concentric large-signal case.}
\label{tab:params}
\begin{ruledtabular}
\begin{tabular}{lcc}
 & Small-signal & Large-signal \\
\colrule
Geometry parameter $C$
  & $0.05$ & $1.95$ \\
Mirror radius $R_0$ (m)
  & $1.0$ & $1.0$ \\
Single-pass length $L_\mathrm{cav}$ (m)
  & $0.05$ & $1.95$ \\
Beam waist $w_0$ ($\mu$m)
  & $\sim226$ & $\sim226$ \\
Path-averaging factor $F(C)$
  & $0.9916$ & $0.2261$ \\
Argon pressure $p_\mathrm{Ar}$ (bar)
  & $5$ & $5$ \\
Gas temperature $T$ (K)
  & $293$ & $293$ \\
Central wavelength $\lambda_0$ (nm)
  & $1030$ & $1030$ \\
Pulse duration $\tau_\mathrm{FWHM}$ (ps)
  & $7.07$ & $7.07$ \\
Pulse energy $E_\mathrm{in}$ (mJ)
  & $22$ & $10$ \\
Pump peak power $P_\mathrm{p,in}$ (GW)
  & $2.92$ & $1.33$ \\
Number of passes $J$
  & $90$ & $70$ \\
Mirror power reflectivity $\mathcal{R}$
  & $0.995$ & $0.995$ \\
\end{tabular}
\end{ruledtabular}
\end{table}

In the simulations, the nonlinear sum in Eq.~\eqref{eq:mmgnlse} is
restricted to the pump self-overlap $S_{0,0,0,0}$ and the diagonal
pump--sideband overlaps $S_{p,0,0,p}$. The nonlinear evolution of the
fundamental mode is therefore
\begin{equation}
  \begin{split}
  \left.
  \frac{\partial A_0}{\partial z}
  \right|_{\mathrm{NL}}
  &=
  i\gamma_\mathrm{eff}(C)
  S_{0,0,0,0}|A_0|^2A_0
  \\
  &\quad
  +i\gamma_\mathrm{eff}(C)
  \sum_{p=1}^{4}S_{p,0,0,p}
  \left(2|A_p|^2A_0+A_0^{*}A_p^2\right),
  \end{split}
  \label{eq:mmgnlse_pump_truncated}
\end{equation}
whereas each retained higher-order radial mode satisfies
\begin{equation}
  \left.
  \frac{\partial A_p}{\partial z}
  \right|_{\mathrm{NL}}
  =
  i\gamma_\mathrm{eff}(C)S_{p,0,0,p}
  \left(2|A_0|^2A_p+A_0^2A_p^{*}\right).
  \label{eq:mmgnlse_sideband_truncated}
\end{equation}
Here $p=1,\ldots,4$. Equations~\eqref{eq:mmgnlse_pump_truncated} and
\eqref{eq:mmgnlse_sideband_truncated} retain pump SPM, pump-induced
XPM of every sideband mode, phase-sensitive FWM, and common-pump
back-action. Terms that directly mix two different higher-order radial indices are excluded. Higher-order-mode back-action on the pump is switched off in the small-signal calculations and retained in the large-signal calculations.

The material-dispersion and Gouy-phase operators are applied in the
spectral domain, and the retained nonlinear equations are integrated
with a fourth-order Runge--Kutta method. After each mirror-to-mirror
pass, every modal amplitude is multiplied by $\sqrt{\mathcal{R}}$, with $\mathcal{R}=0.995$. The nonlinear coefficient is the path-averaged value $\gamma_\mathrm{eff}(C)=\gamma_0F(C)$ from Eq.~\eqref{eq:gamma_eff}. The intra-pass variation of $w(\zeta)$ is
therefore not propagated explicitly. The small-signal spectral-evolution and modal-gain calculations use $2^{14}$ temporal points over windows of $40$ and $30$~ps, respectively. The large-signal calculations use $2^{15}$ temporal points over a $100$-ps window. The input temporal intensity profile is
\begin{equation}
  P(t)=P_\mathrm{p,in}
  \exp\!\left[
    -4\ln 2
    \left(\frac{t}{\tau_\mathrm{FWHM}}\right)^2
  \right],
  \label{eq:input_pulse}
\end{equation}
where $P_\mathrm{p,in}=2\sqrt{\ln 2/\pi} E_\mathrm{in}/\tau_\mathrm{FWHM}$ is the peak power.

\begin{acknowledgments}
\noindent This work was partially supported by National Key R$\&$D Program of China (2024YFB3613503) and National Natural Science Foundation of China (62275015 and 62205015).
\end{acknowledgments}

\section*{Data Availability}

The numerical data and custom simulation code that support the findings
of this study are available from the corresponding author upon
reasonable request.

\bibliography{Reference}
 
\end{document}